\documentclass[reprint,aps,prr,nobibnotes,superscriptaddress]{revtex4-2}

\usepackage{siunitx,braket,graphicx,xcolor,amsmath,amssymb,array}
\usepackage{ulem}

\newcommand{\avg}[1]{\left\langle #1 \right\rangle}
\newcommand{\Tr}[1]{\mathrm{Tr}\left[ #1 \right]}
\newcommand{\ZPF}{\mathcal{M}_{\rm ZPF}}
\newcommand{\nY}{\mu_{\rm ref}}
\newcommand{\opabs}{\alpha_{\rm abs}}
\newcommand{\vecOp}[1]{\hat{\boldsymbol{#1}}}

\begin{document}
\title{Quantum tomography of magnons using Brillouin light scattering}
\author{Sanchar Sharma}
\affiliation{Institut f{\"u}r Theoretische Festk{\"o}rperphysik, RWTH Aachen University, 52056 Aachen, Germany}
\author{Silvia Viola Kusminskiy}
\affiliation{Institut f{\"u}r Theoretische Festk{\"o}rperphysik, RWTH Aachen University, 52056 Aachen, Germany}
\affiliation{Max Planck Institute for the Science of Light, 91058 Erlangen, Germany}
\author{Victor A.S.V. Bittencourt}
\affiliation{ISIS (UMR 7006), Universit{\'e} de Strasbourg, 67000 Strasbourg, France}

\begin{abstract}
Quantum magnonics, an emerging field focusing on the study of magnons for quantum applications, requires precise measurement methods capable of resolving single magnons. Existing techniques introduce additional dissipation channels and are not apt for magnets in free space. Brillouin light scattering (BLS) is a well-established technique for probing the magnetization known for its high sensitivity and temporal resolution. The coupling between magnons and photons is controlled by a laser input, so it can be switched off when a measurement is not needed. In this article, we theoretically investigate the efficacy of BLS for quantum tomography of magnons. We model a finite optomagnonic waveguide, including the optical noise added by the dielectric, to calculate the signal-to-noise ratio (SNR). We find that the SNR is typically low due to a small magneto-optical coupling; nevertheless, it can be significantly enhanced by injecting squeezed vacuum into the waveguide. We reconstruct the density matrix of the magnons from the statistics of the output photons using a maximum likelihood estimate. The classical component of a magnon state, defined as the regions of positive Wigner function, can be reconstructed with a high accuracy while the non-classical component necessitates either a higher SNR or a larger dataset. The latter requires more compact data structures and advanced algorithms for post-processing. The SNR is limited partially by the input laser power that can be increased by designing the optomagnonic cavity with a heat sink.
\end{abstract}

\maketitle

Magnons, the quanta of spin-waves, are promising candidates for non-silicon based devices for wave or quantum computing~\cite{Roadmap}. They couple to a variety of excitations such as optical photons~\cite{OptMag_Osada, OptMag_Zhang, OptMag_Silvia, OMag_WG_Liu, OptMag_Sanch, OptMag_James, OMag_WG_Zhu}, microwaves~\cite{MagMW_Th, MagMW_Exp1, MagMW_Exp2}, phonons~\cite{MagPh1, MagPh2}, and spin centers~\cite{MagNV_Luka, MagNV_Toeno, MagNV_Bertelli, MagNV_Carlos} making them versatile transducers. In particular, magnonic systems based on Yttrium Iron Garnet (YIG)~\cite{YIG_Saga} are interesting owing partly to their low dissipation, even in nanoscale samples~\cite{Schmidt_Rev, YIGWaveguide}. Furthermore, since the magnon frequency~\cite{MagMW_Exp1, MagMW_Exp2} and non-linearities~\cite{YIGWaveguide, Mehrdad_NL} are externally tunable, they are candidates for scalable bosonic qubits  suitable for quantum error correction \cite{gottesman_encoding_2001, grimsmo_quantum_2021, terhal_towards_2020, cai_bosonic_2021}.

Using magnons for quantum applications requires the ability to generate and measure non-classical states of the magnetization. The first states of this kind were recently experimentally demonstrated~\cite{Xu_MagFock}. Theory proposals for state generation in magnons includes both heralded measurement techniques \cite{bittencourt_magnon_2019, sun_remote_2021, sharma_spin_2021}, and deterministic protocols \cite{sharma_protocol_2022, kounalakis_analog_2022}. Nevertheless, quantum measurements of magnon states is less explored. Classically, magnons can be probed using microwaves via ferromagnetic resonance (FMR) experiments~\cite{StanPrabh}, NV centers~\cite{MagNV_Toeno,MagNV_Bertelli}, electric currents via spin pumping~\cite{SMR_Nakayama, Ulloa_NlST, hioki_2021_state_tomography}, and optical light via Brillouin light scattering (BLS)~\cite{BLS_Rev1, BLS_Rev2}. Microwave cavities can mediate a coupling between the magnons and a superconducting transmon, which has been demonstrated to perform magnon tomography~\cite{lachance-quirion_entanglement-based_2020, tabuchi_quantum_2016, wolski_dissipation-based_2020, Xu_MagFock}. Such a measurement would not be suitable for applications which require a magnet to be in free space, such as magnetic field sensing. Spin pumping into a spin-Hall material, such as Pt, can be used for performing `classical tomography' of magnons~\cite{hioki_2021_state_tomography}, however it is in principle not suitable for measuring magnon coherence because of the dissipative nature of spin-to-charge converters.

BLS is one of the most sensitive probes of the magnetization with a high spatial and temporal resolution~\cite{BLS_Rev2}, which has allowed for detecting and mapping spin-wave eigenmodes \cite{vogt_2011_opticaldetection} and for investigating non-linear processes \cite{schultheiss_2019_excitationof} in microdisks. BLS has also been recently used to probe spin-wave time refraction in strips \cite{schultheiss_2021_timerefraction}, and to observe high-momentum magnons \cite{wojewoda2023}. Unlike methods employing microwaves or spin pumping, BLS does not lead to extra dissipation because the coupling between the magnons and the photons can be switched on and off using an external laser. When compared to other methods, BLS might then be a useful tool to also probe features of quantum states of magnons. 

\begin{figure}
    \centering
    \includegraphics[width=\columnwidth]{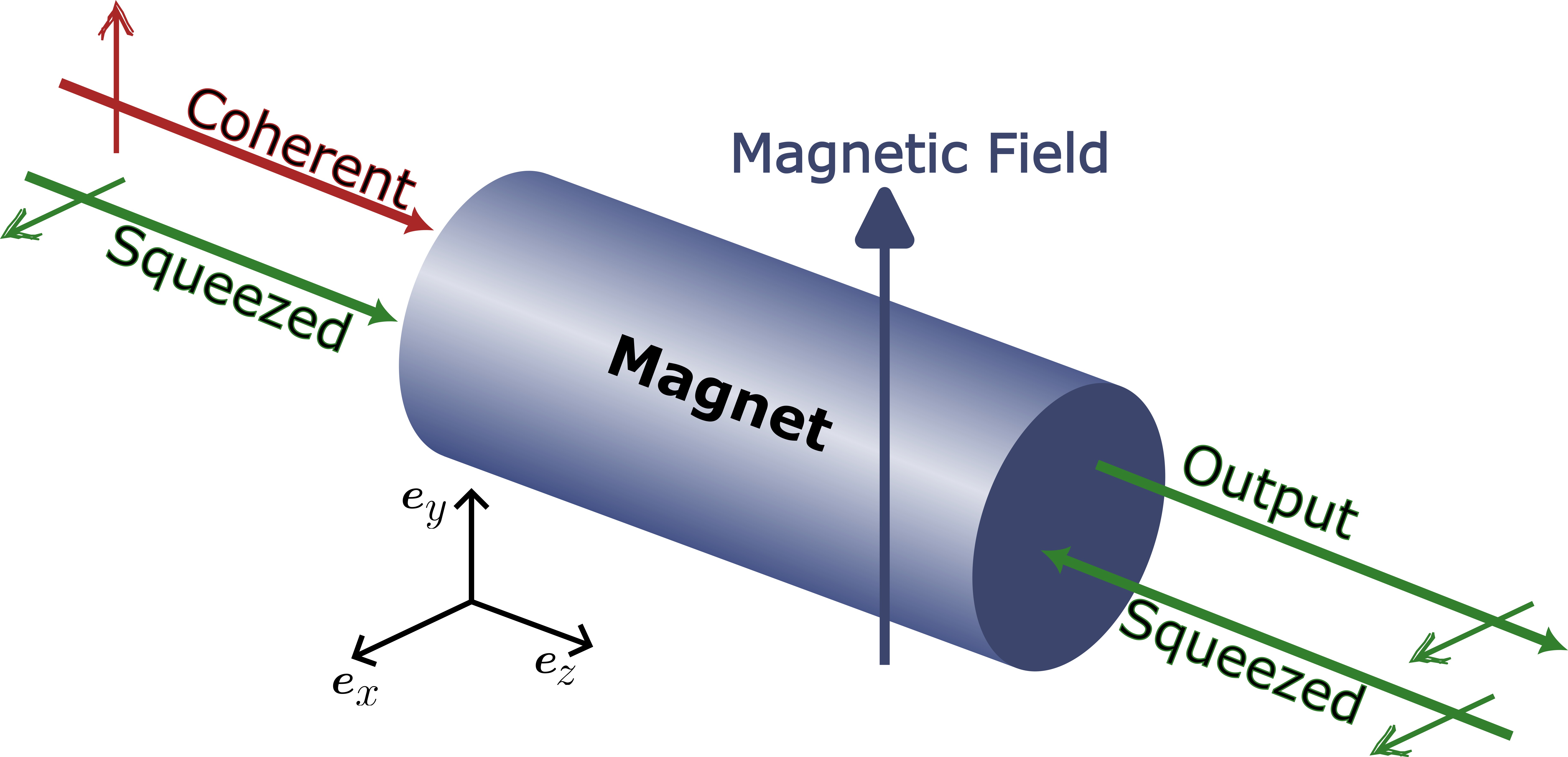}
    \caption{Envisioned setup consisting of a waveguide made of a magnetic material. The red incoming line denotes a large coherent optical input in the $\boldsymbol{e}_y$-polarization. The green incoming lines denote squeezed vacuum in the $\boldsymbol{e}_x$-polarization while the outgoing line is the output to be measured. A static magnetic field saturates the magnetization along $\boldsymbol{e}_y$.}
    \label{fig:System}
\end{figure}

In this work, we theoretically propose and evaluate a method for BLS-based tomography of magnon quantum states. For simplicity and concreteness, we consider an optomagnonic waveguide~\cite{OMag_WG_Liu, OMag_WG_Zhu} made of a magnetic dielectric as shown in Fig.~\ref{fig:System}, including the dielectric losses and added noise. This geometry offers moreover an optimal overlap between the photons and the magnons. The results are qualitatively general and hold for other geometries such as spheres. A smaller overlap between the photons and the magnons mode profiles leads however to a decrease in the signal. 

Input light propagating through the waveguide can inleastically scatter via magnons into the perpendicular polarization, see Fig.~\ref{fig:System}, such that the output light of the waveguide has imprints of the magnon state. The homodyne data of the output light can then be used to estimate the magnon's density matrix using a maximum likelihood estimate (MLE). To reduce the output noise, we consider a squeezed thermal input in the polarization into which light is scattered. Such a tomographic reconstruction is limited by a trade-off between the strength of the magneto-optical coupling and optical dissipation in the material, which typically go hand-in-hand~\cite{wettling_magneto-optics_1976}. We numerically characterize the fidelity of the tomography under different signal-to-noise ratios for a classical (squeezed coherent) state and a non-classical (cat) state of the magnetization. To this end, we simulate experimental data by analytically calculating the output optical probability density for a given magnon state.

In Sec.~\ref{sec:Recons}, we discuss and evaluate an MLE-based method of estimating the magnon's density matrix using the statistics of the output photons for a general SNR and input squeezing. In Sec.~\ref{sec:WG_ana}, we analyze our system in Fig.~\ref{fig:System} to find the output light by solving the electromagnetic Hamiltonian which includes BLS and optical damping within the magnet. We conclude in Sec.~\ref{sec:Conclusion}.

\section{Reconstruction of a magnon state} \label{sec:Recons}

In an optomagnonic waveguide, see Fig.~\ref{fig:System}, a large $\boldsymbol{e}_y$-polarized input pulse facilitates the exchange of quanta from magnons to $\boldsymbol{e}_x$-polarized photons. The $\boldsymbol{e}_x$-polarized output is affected by external noise and by impurity noise within the magnet. In this section, we discuss the tomographic reconstruction of a magnon state from a noisy output optical signal for a general SNR and input squeezing.

The general procedure is described in Sec.~\ref{Recons_sub:Eval}. The fidelity of reconstruction is discussed for a pure Gaussian state in Sec.~\ref{Recons_sub:Gauss} and a cat state in Sec.~\ref{Recons_sub:Cat}. To understand the reconstruction fidelity, we discuss the output optical probability distribution explicitly in Sec.~\ref{Recons_sub:Prob}.

\subsection{Evaluation Procedure} \label{Recons_sub:Eval}

The output photons are a mixture of optical noise and magnons, such that the output annihilation operator $\hat{a}_{\rm out}$ is given by
\begin{equation}
    \hat{a}_{\rm out} = \cos\theta\hat{\eta}_{\rm{out}}+\sin\theta\hat{m}, \label{eq:out}
\end{equation}
where  $\hat{\eta}_{\rm out}$ is the noisy output if there were no coupling to magnons, and $\hat{m}$ is the annihilation operator of magnons. The angle $\theta$ is a sample-dependent quantity of magneto-optical origin, whose formula we discuss in detail in Sec.~\ref{sec:Numbs}, which gives the signal to noise ratio (SNR)  $\tan \theta$. As discussed there, we expect $\theta\sim 0.2$ for an optimally coupled optomagnonic cavity with no heat sink. For evaluating the reconstruction procedure, we consider three cases, $\theta/\pi\in\{0.02,0.25,0.45\}$, in increasing order of SNR $\tan\theta\in\{0.06,1,6.3\}$. For completeness, we discuss the case of higher $\theta$ assuming future advancements in cavity and material design as well as lower $\theta$ that can occur due to experimental imperfections.

\begin{figure}
\includegraphics[width = \columnwidth]{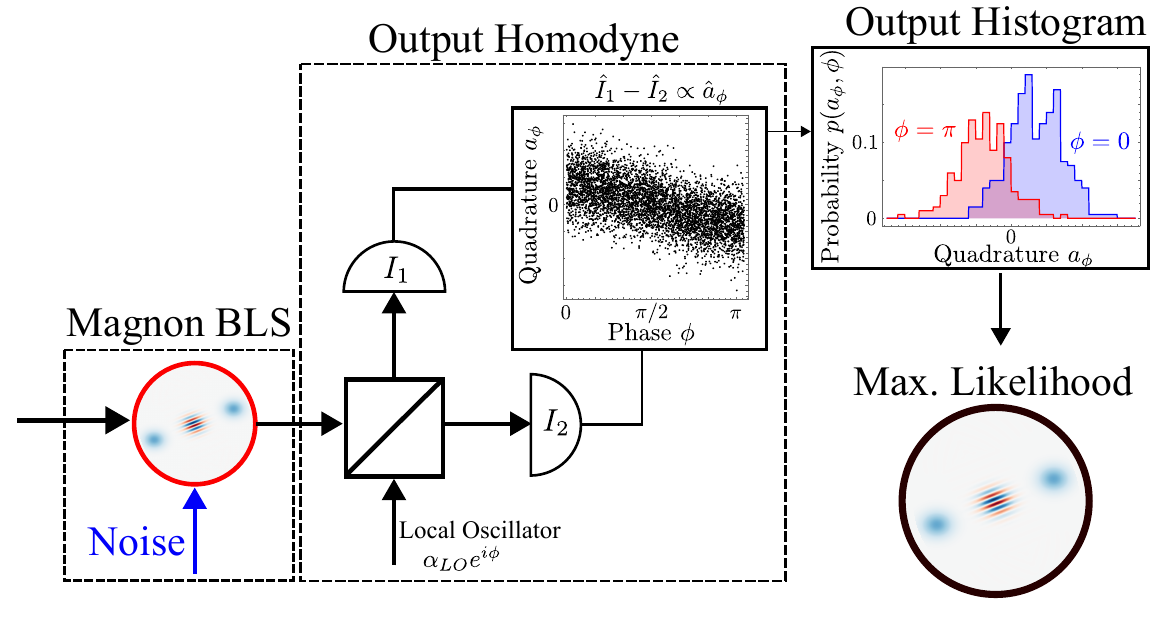}
\caption{Schematic representation of the magnon tomography. The input light is scattered via magnons, acquiring a noisy signal which encodes the magnon state. A homodyne measurement on the output photons is performed by mixing it with a local oscillator with an amplitude $\alpha_{LO}$, providing amplification, and phase $\phi$, setting the phase of the quadrature to be measured, $\hat{a}_{\phi}$ as defined in Eq.~(\ref{def:aphi}). A set of data corresponding to the measured values of $\hat{a}_{\phi}$ for a given $\phi$ is acquired by subtracting the photocurrents $\hat{I}_{1}$ and $\hat{I}_{2}$. Such a set is then used in the maximum likelihood procedure explained in the text, giving an estimate of the magnon state. Data and probability histogram depicted in the figure are illustrative. \label{fig:Scheme}}
\end{figure}

The homodyne data consists of measurements of the quadratures of the output $\hat{a}_{\rm{out}}$,
\begin{equation}
    \hat{a}_{\phi} = \hat{a}_{\rm out} e^{-i\phi} + \hat{a}_{\rm out}^{\dagger} e^{i\phi}, \label{def:aphi}
\end{equation}
by varying $\phi$ uniformly in the range $[0,\pi]$. The phase $\phi$ is set by a local oscillator mixed with the output signal via a beam splitter, see Fig.~\ref{fig:Scheme}. A set of experimental measurements can be written as $\{(a_{i},\phi_{i})\}$ for $i\in\{1,\dots,N\}$ where $N$ is the number of measured samples, $\phi_i$ is the chosen $\phi$ for the $i$th measurement, and $a_i$ is the measured value therein. Since $\hat{a}_{\rm{out}}$ includes contributions due to the magnons, the output homodyne data can be used to reconstruct the magnon state as we now discuss.

For a given $\phi$, the random variable $\hat{a}_{\phi}$ follows a probability density $p_a(a,\phi)$. We can write $p_a$ in terms of the magnon's density matrix $\hat{\rho}_m$ as
\begin{equation}
    p_a(a,\phi) = \Tr{\hat{\rho}_m \hat{P}(a,\phi)}, \label{eq:prob_a}
\end{equation}
where the operator $\hat{P}$ is
\begin{equation}
    \hat{P}(a,\phi) = \frac{1}{\cos\theta} p_{\eta}\left(\frac{a-\sin\theta\hat{m}_{\phi}}{\cos\theta},\phi\right).\label{def:Proj}
\end{equation}
We show the detailed derivation of Eq.~\eqref{eq:prob_a} in Sec.~\ref{Recons_sub:Prob}. Here $p_{\eta}(\eta,\phi)$ is the probability density of the noise quadrature $\hat{\eta}_{\phi}$, where both $\hat{m}_{\phi}$ and $\hat{\eta}_{\phi}$ are defined analogous to $\hat{a}_{\phi}$. Under equilibrium conditions, $p_{\eta}$ would correspond to the vacuum fluctuations. However, we can significantly decrease the noise by injecting squeezed vacuum, cf. Fig.~\ref{fig:System}, such that 
\begin{equation}
    p_{\eta}(\eta,\phi) = \frac{1}{\sqrt{2\pi}\sigma_s}\exp\left[\frac{-\eta^2}{2\sigma_s^2} \right], \label{def:noise_prob}
\end{equation}
where $\sigma_s^2 \le 1$ is the variance of the squeezed noise quadrature. As $\phi$ is tuned externally for each data point, the squeezing can be adjusted such that the quadrature $\hat{\eta}_{\phi}$ is squeezed. We show in Sec.~\ref{sec:WG_ana} that $\sigma_s \approx e^{-r_{\rm in}}$ where $r_{\rm in}$ is the squeezing parameter of the input, when the output is slightly off-resonance with the cavity [see Eq.~(\ref{eq:ns:var})]. This implies that the squeezing is not significantly affected by the optical impurities in the magnet. 

When either $\theta\rightarrow\pi/2$ (infinite SNR) or $\sigma_s \rightarrow 0$ (no noise), we get $\hat{P}\rightarrow \delta(\sin\theta \hat{m}_{\phi} - a)$, implying that the probability of measuring $\hat{a}_{\phi}$ to be $a$ is exactly the probability of measuring $\hat{m}_{\phi}$ to be $a/\sin\theta$. In this case, the magnon probability density can be directly reconstructed using the optical probability density. In general, we can interpret Eq.~\eqref{eq:prob_a} as follows: the probability of measuring $\hat{a}_{\phi}$ to be $a$ gives an estimate of $\hat{m}_{\phi}$ being close to $a/\sin\theta$ with an error of $\sigma_s \cos\theta$, as expected from Eq.~\eqref{eq:out}.

Eq.~(\ref{eq:prob_a}) can, in principle, be inverted exactly to find $\hat{\rho}_m$ in terms of the probability density of the observed optical data. However, it is known from classical estimation theory~\cite{Deconv} that an exact inversion amplifies noise exponentially, making its numerical implementation infeasible. We then resort to a statistical procedure using the maximum likelihood principle, i.e. given a set of data, say $(a_{i},\phi_{i})$ for $i\in\{1,\dots,N\}$, we want to find a magnon's density matrix $\hat{\rho}_m$ that maximizes the probability of observing this dataset, i.e. maximize $\prod p_a(a_i,\phi_i)$. The noiseless case ($\theta = \pi/2$ in Eq.~(\ref{eq:out})) has been solved in the context of optical tomography \cite{lvovsky_continuous-variable_2009,lvovsky_iterative_2004,fiurasek_maximum-likelihood_2001,Hradil_QSE}. Analogous to these works, we show in App.~\ref{app:ML} that the optimal density matrix can be found via the recursion relation,
\begin{equation}
    \hat{\rho}_{k+1}=\frac{\hat{Z}(\hat{\rho}_{k})\hat{\rho}_{k} + \hat{\rho}_{k} \hat{Z}(\hat{\rho}_{k})}{2}, \label{recursion}
\end{equation}
starting from an arbitrary initial guess $\hat{\rho}_0$, where
\begin{equation}
    \hat{Z}(\hat{\rho})=\frac{1}{N}\sum_{i=1}^{N}\frac{\hat{P}(a_{i},\phi_{i})}{{\rm Tr}\left[\hat{\rho}\hat{P}(a_{i},\phi_{i})\right]}.
\end{equation}
Note that if the initial guess is a valid density matrix, i.e. $\hat{\rho}_{0}=\hat{\rho}_{0}^{\dagger}$, $\hat{\rho}_{0}\ge0$, and ${\rm Tr}\left[\hat{\rho}_{0}\right]=1$, each $\hat{\rho}_{k}$ is also a valid density matrix. The convergence is guaranteed by the Banach's fixed point theorem.

To evaluate the fidelity of reconstruction for a given magnon target state $\Ket{\Psi_{\rm tar}}$, we simulate an experimental data set by generating $N=10^4$ optical data points as follows. We analytically derive $p_a(a,\phi)$ corresponding to a given magnon signal amplitude $\theta$ and magnon state $\Ket{\Psi_{\rm tar}}$ using Eq.~\eqref{eq:prob_a}. We choose a $\phi\in[0,\pi)$ uniformly and sample $a$ using the probability distribution $p_a(a,\phi)$. The generated data is then processed using MLE, Eq.~(\ref{recursion}), to find the prediction, giving a workflow
\begin{equation}
    \Ket{\Psi_{\rm tar}} \rightarrow \{(a_i,\phi_i)\}_{i=1}^{10^4} \rightarrow \hat{\rho}_{\rm pred}.
\end{equation}
We then plot the Wigner functions of the target and the predicted states along with calculating the fidelity $F = \sqrt{\Braket{ \Psi_{\rm tar} | \hat{\rho}_{\rm pred} | \Psi_{\rm tar} } }$. The number of simulated data points $N$ is chosen to be small for computational reasons.

\subsection{Gaussian state} \label{Recons_sub:Gauss}

First, we evaluate the reconstruction of a squeezed coherent state defined as, 
\begin{equation}
    \Ket{\Psi_{\rm coh}} = \hat{D}(\alpha_s) \hat{S}(r_s,\psi_s) \Ket{0}. \label{def:SqCoh}
\end{equation}
where the operators $\hat{D}(\alpha_s)$ and $\hat{S}(r_s,\psi_s)$ are respectively the displacement and the squeezing operator of the magnons (see App.~\ref{app:StateGallery} for the definitions). Magnon squeezing can be generated by, for example, geometric anisotropy~\cite{Akash_Sq,sharma_spin_2021} and displacements can be generated by a microwave excitation. We choose an arbitrary set of parameters $\alpha_s = 1.8-2.4i$, $e^{r_{s}}=1.5$, and $\psi_{s}=2.7$.

\begin{figure}
\includegraphics[width = \columnwidth]{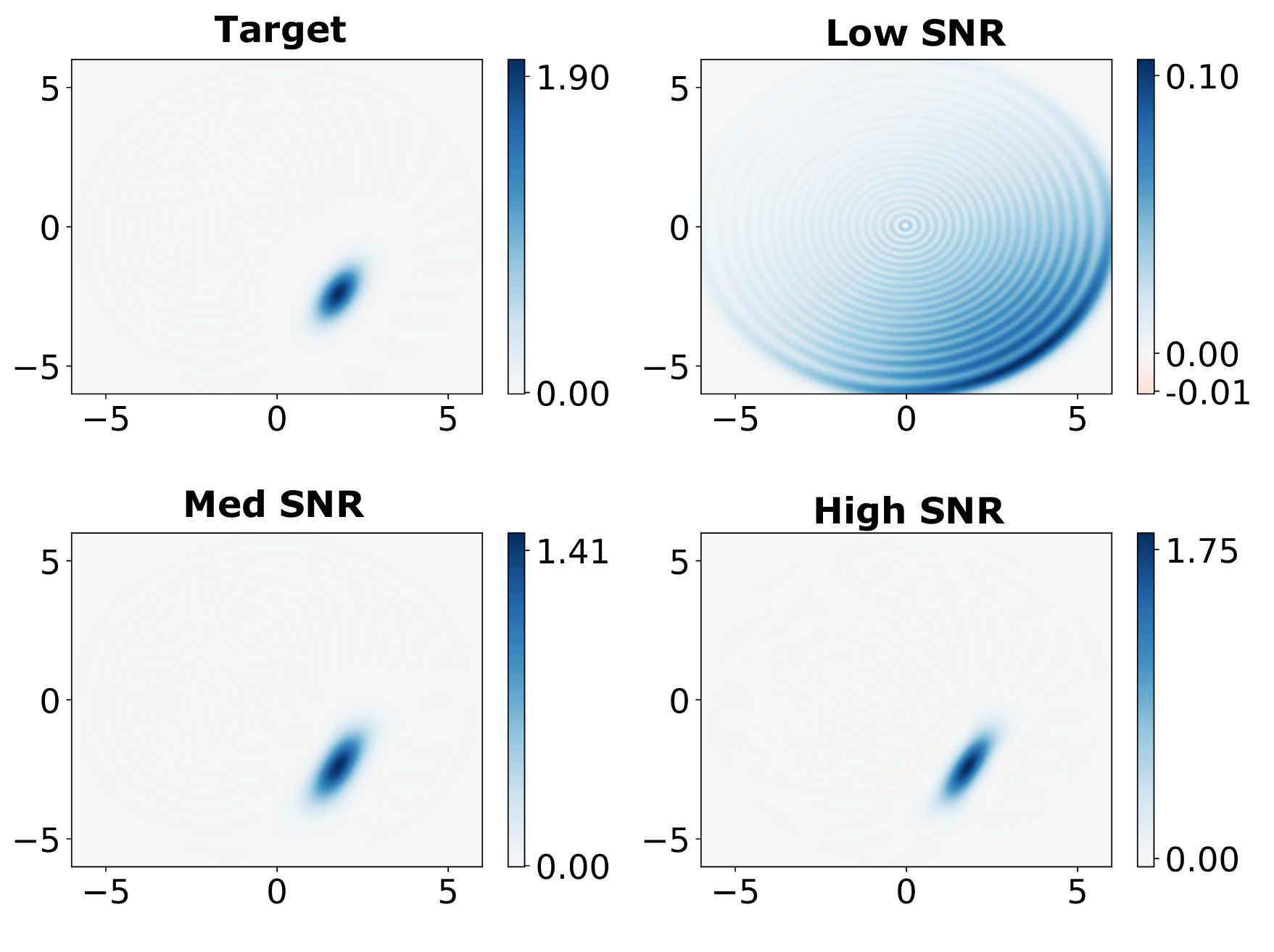}
\caption{The Wigner functions of the target state $\Ket{\Psi_{\rm coh}}$, and the reconstructed outputs for $\theta = 0.02\pi$ (Low SNR), $\theta = \pi/4$ (Med SNR), and $\theta = 0.45\pi$ (High SNR), without the enhancement due to the squeezing of optical input. The fidelities of reconstruction in the increasing order of SNR are $0.21$, $0.92$, and $0.96$.  \label{fig:Coh_woSq}}
\end{figure}

In Fig. \ref{fig:Coh_woSq}, we discuss the results when there is no input optical squeezing, i.e. $\sigma_s = 1$ in Eq.~\eqref{def:noise_prob}. The Wigner functions of the target state and the reconstructed outputs for the three cases of $\theta$ are shown. As expected, the high SNR case gives a high fidelity of reconstruction. Perhaps slightly surprisingly, the $\text{SNR}=1$ case also gives nearly the same fidelity, although it underestimates the degree of squeezing. For the low SNR case, the prediction severely overestimates the number of magnons. This can be qualitatively understood as follows. For $\theta\ll 1$, there is a very small probability of observing a photon. So, each observed photon is taken by the algorithm as a strong indication of the presence of a high number of magnons, essentially amplifying the noise. The predicted $\langle \hat{m}\rangle =1.5-1.9i$ is smaller in magnitude than the target displacement $\alpha_s$, although they both have almost the same angle with an error of less than $\SI{2}{\degree}$ (note that the number of magnons is much larger than $|\langle \hat{m} \rangle|^2$). Thus, even for a very low SNR, the magnon phase can be predicted with a high accuracy which is useful for measuring weak magnetic fields~\cite{Kimball_2016}.

\begin{figure}
\includegraphics[width = \columnwidth]{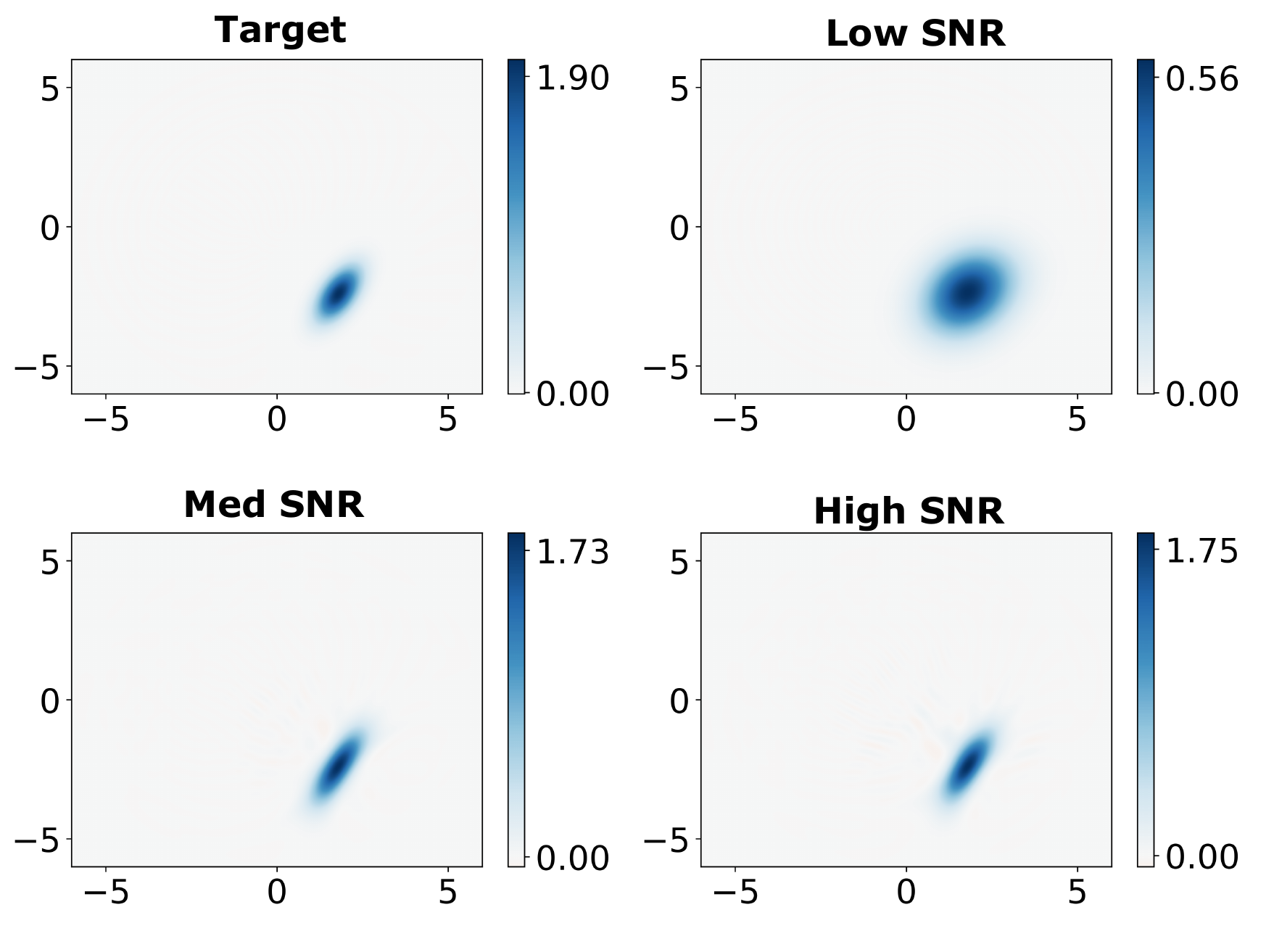}
\caption{Similar plot as in Fig.~\ref{fig:Coh_woSq} but with squeezed input with the standard deviation in the squeezed quadrature being $\sigma_s^{-1} = 5$ times smaller than its vacuum value. The fidelities of reconstruction in increasing order are $0.67$, $0.96$, $0.97$.
\label{fig:Coh_WithSq}}
\end{figure}

Now, we consider the case with squeezed input with a realistically achievable parameter $\sigma_s^{-1}=5$~\cite{Andersen_SqLight,Eberle_SqLight,Mehmet_SqLight}, that should improve the reconstruction fidelities. As we see in Fig. \ref{fig:Coh_WithSq}, even for the low SNR case, we get the correct $\left\langle \hat{m}\right\rangle $ although not the squeezing parameters. Medium and high SNR cases are indistinguishable. Here, the reason for non-unity fidelity is the choice of a smaller Hilbert space cut-off for computational reasons. Note that the cut-off is not given by the magnon state but rather by a faithful representation of the operator $\hat{P}$ [see Eqs.~(\ref{def:Proj})] which approximates a $\delta$-function for $\tan\theta/\sigma_s \gg 1$.

\subsection{Cat state} \label{Recons_sub:Cat}
The second state we consider is a cat state,
\begin{equation}
    \Ket{\Psi_{\rm cat}} = \mathcal{N} \frac{\Ket{\alpha_c} + e^{i\psi_c} \Ket{-\alpha_c}}{\sqrt{2}}, \label{def:Cat}
\end{equation}
where $\mathcal{N}$ is a normalization constant, and $\Ket{\pm \alpha_c} = \hat{D}(\pm \alpha_c)\Ket{0}$. Magnon cat states can be generated using heralding protocols assisted by microwaves or optical photons~\cite{sharma_spin_2021, sun_remote_2021}, as well as deterministically~\cite{sharma_protocol_2022,kounalakis_analog_2022}. We arbitrarily choose $\alpha_c = 2.7+1.3j$ and $\psi_c = -1.7$. For later comparison, we define a 50:50 classical mixture of coherent states which can be thought of as the `classical component' of $\Ket{\Psi_{\rm cat}}$,
\begin{equation}
    \hat{\rho}_{\rm cl} = \frac{\Ket{\alpha_c}\Bra{\alpha_c} + \Ket{-\alpha_c}\Bra{-\alpha_c}}{2}. \label{def:rho_cl}
\end{equation}

\begin{figure}
\includegraphics[width = \columnwidth]{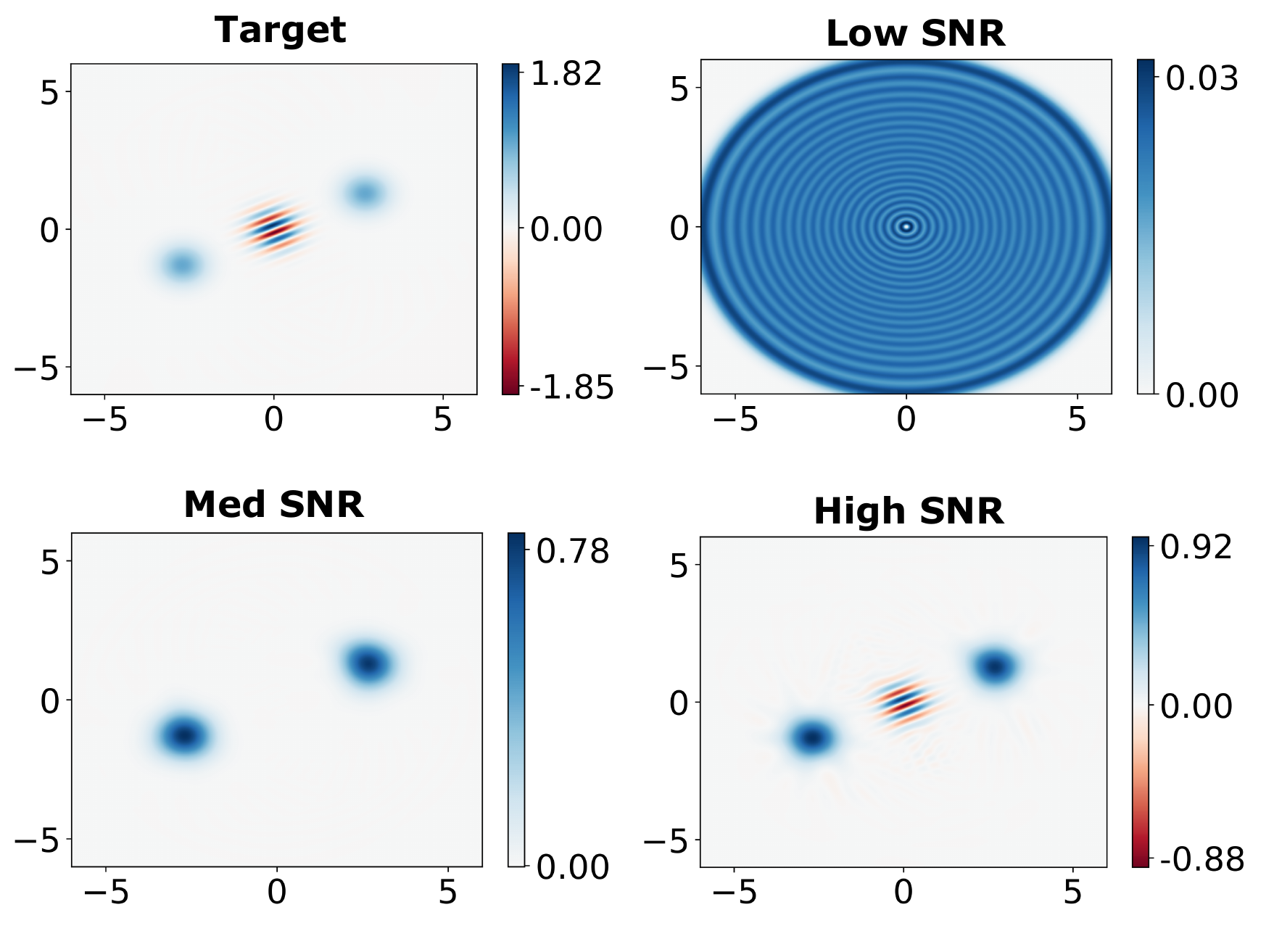}
\caption{Wigner functions for a target cat state $\Ket{\Psi_{\rm cat}}$ for different SNRs similar to Fig.~\ref{fig:Coh_woSq}. The fidelities of reconstruction in increasing order are $0.16$, $0.67$, and $0.85$. For the case of medium SNR, the reconstruction has a $0.95$ fidelity with the classical mixture $\hat{\rho}_{\rm cl}$. In the high SNR case, the fidelity is limited by the small number of samples considered here. \label{fig:Cat_woSq}}
\end{figure}

\begin{figure}
\includegraphics[width = \columnwidth]{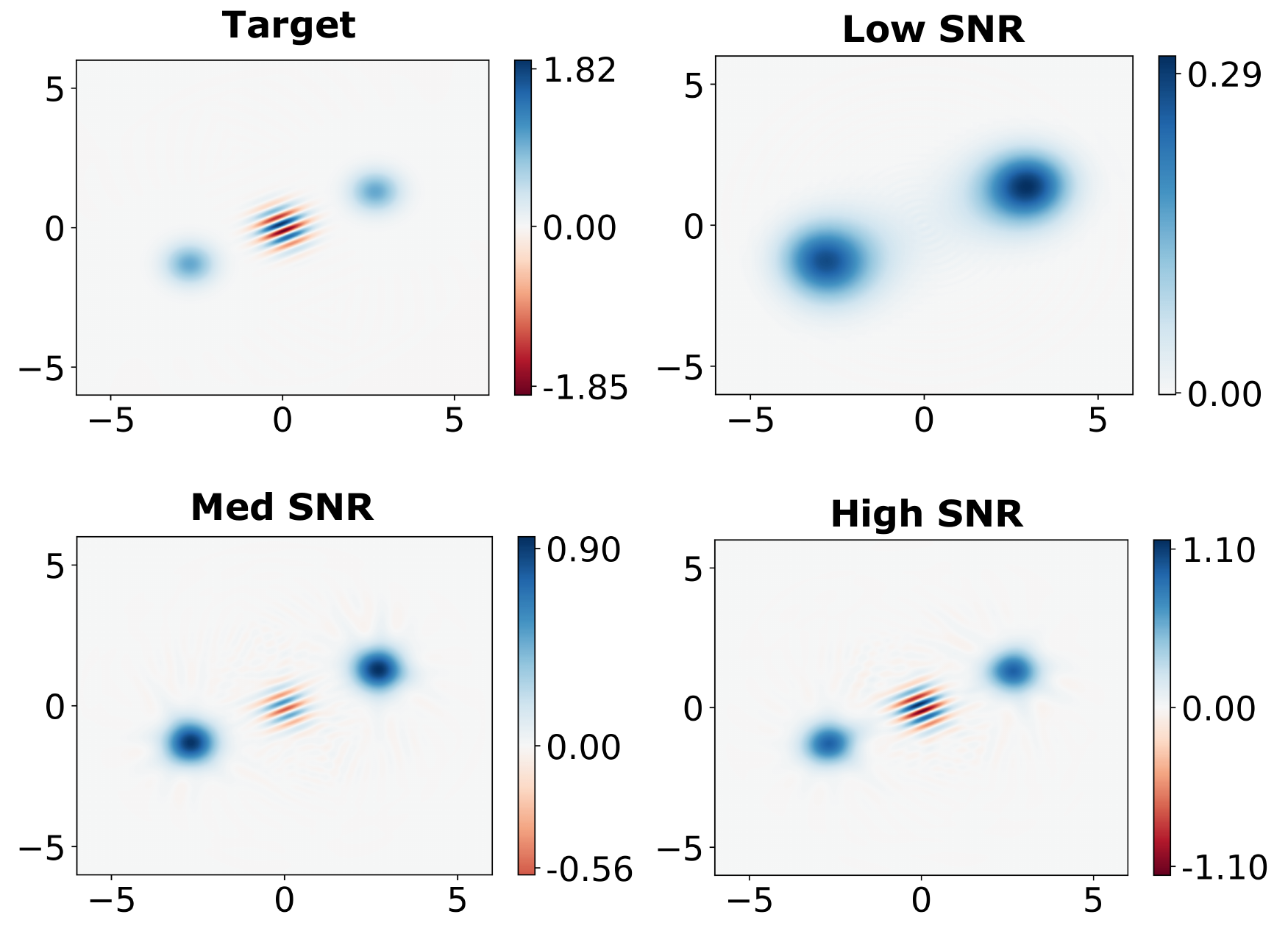}
\caption{Similar plot as in Fig.~\ref{fig:Cat_woSq} but with squeezed input with parameter $\sigma_s^{-1} = 5$. The fidelities of reconstruction in increasing order are $0.46$, $0.79$, $0.88$. For the case of low SNR, the reconstruction has a $0.66$ fidelity with the classical mixture $\hat{\rho}_{\rm cl}$. \label{fig:Cat_WithSq}}
\end{figure}

In Fig. \ref{fig:Cat_woSq}, we discuss the case with no optical squeezing, $\sigma_s = 1$ in Eq.~\eqref{def:noise_prob}. The high SNR case gives a moderate fidelity of reconstruction, while maintaining the salient features of a cat state, in particular the interference fringes. Here, the fidelity is limited by the number of data points, $N=10^4$ in our case. The low SNR case gives the identity matrix again amplifying the noise to a level that no magnon information is retrievable. 

The medium SNR case predicts the closest classical approximation to the cat state, i.e. $\hat{\rho}_{\rm cl}$ as defined in Eq.~(\ref{def:rho_cl}). This can be explained by comparing the output probability distributions for the cat state $\Ket{\Psi_{\rm cat}}$ and the classical mixture $\hat{\rho}_{\rm cl}$ (see the discussion below Eq.~\eqref{pcat:int}). We find that these output statistics are approximately same when $\tan\theta/\sigma_s < |\alpha_c|$. Very subtle statistical differences can be detected with a larger data set, although that will increase the computational time. Thus, more efficient computational algorithms can further improve the reconstruction fidelity.

In Fig. \ref{fig:Cat_WithSq}, we discuss the case with input optical squeezing. The low SNR case now predicts the correct peaks of the distribution. However, it misses the coherence and predicts a larger noise than present in the state. This is tantamount to the amplification of noise as observed before with much less severity. With increasing SNR, the prediction becomes better. In particular, we notice that at a medium SNR, the tomographic reconstruction already captures the interference fringes which are absent in Fig.~\ref{fig:Cat_woSq}. The fidelity improves as the number of data points are increased (not shown here for brevity), showing again that quantum information typically requires observing more subtle changes in the probability distribution.

\subsection{Analytical probability distributions} \label{Recons_sub:Prob}
Here, we discuss the output optical probability distribution for the two cases considered in Secs.~\ref{Recons_sub:Gauss} and \ref{Recons_sub:Cat}. We find the effect of noise and dissipation on the optical data, and in particular, show analytically that quantum coherence in a cat state is lost even when the information about the peak amplitudes is preserved, as found in Figs.~(\ref{fig:Cat_woSq},\ref{fig:Cat_WithSq}). The salient features of deriving the output optical probability distribution, $p_a$, are written here with intermediate steps in App.~\ref{app:StateGallery}.

For a given $\phi$, the magnon quadratures $\hat{m}_{\phi} = \hat{m}e^{-i\phi} + \hat{m}^{\dagger} e^{i\phi}$ have a probability distribution $p_m(m,\phi)$ defined as the unique distribution that satisfies,
\begin{equation}
    \avg{\hat{f}(\hat{m}_{\phi})} =\int da\ f(m)p_m(m,\phi),\label{def:mag_prob}
\end{equation}
for any function $f$. For a given magnon density matrix $\hat{\rho}_m$, we can calculate the magnon probability distribution formally by putting $f(m) = \delta(m-m_0)$,
\begin{equation}
    p_m(m_0,\phi) = \Tr{\hat{\rho}_m \hat{\delta}(\hat{m}_{\phi} - m_0)}. \label{prob_m:delta}
\end{equation}
Because of the singular nature of the above equation, we discuss a more practical method to calculate $p_m$ below.

We define $p_a$ and $p_{\eta}$ as the probability distributions of $\hat{a}_{\rm out}$ and $\hat{\eta}_{\rm out}$ respectively. Now, we want to find $p_a(a,\phi)$ in terms of the magnon's density matrix. Since the random variables $\hat{\eta}$ and $\hat{m}$  are independent, one can write for any function $g$, 
\begin{equation}
    \avg{\hat{g}(\hat{m}_{\phi},\hat{\eta}_{\phi})} = \int dm dn\  g(m,n) p_m(m,\phi) p_{\eta}(\eta,\phi) \label{def:joint_prob}
\end{equation}
Inserting the relation $\hat{a}_{\phi} = \cos\theta \hat{\eta}_{\phi} + \sin\theta \hat{m}_{\phi}$ (see Eq.~(\ref{eq:out})) into Eq.~(\ref{def:mag_prob}) and then comparing with Eq.~(\ref{def:joint_prob}) for $g(m,\eta) = f(\cos\theta\eta + \sin\theta m)$, we obtain the following relation between the probability distributions,
\begin{multline}
    p_a(a,\phi) = \int da'\ p_{\eta} \left(\cos\theta a-\sin\theta a',\phi\right) \\ \times p_m\left(\sin\theta a+\cos\theta a',\phi\right). \label{prob_a:conv}
\end{multline}
Inserting the relation Eq.~\eqref{prob_m:delta} into the above, we find the dependence of $p_a$ on $\hat{\rho}_m$ as written in Eq.~\eqref{eq:prob_a}.

We can now obtain the output probability statistics for the magnon states $\Ket{\Psi_{\rm coh}}$ and $\Ket{\Psi_{\rm cat}}$. We first calculate the magnon probability distribution by
\begin{equation}
    p_m(m,\phi) = \int \frac{d\beta}{2\pi} \Braket{\Psi | e^{i\beta(m - \hat{m}_{\phi})} | \Psi }.
\end{equation}
To derive this, we put $f(m) = e^{i\beta m}$ for an arbitrary $\beta$ in Eq.~(\ref{def:mag_prob}) to find the Fourier transform of $p_m(m,\phi)$ and then, invert the Fourier transform. 

A similar formula can be used for the noise. Consider the case when the variances of two independent quadratures $\hat{\eta}_{\psi}$ and $\hat{\eta}_{\psi+\pi/2}$ are $\sigma_s$ and $\sigma_b$ respectively, for a chosen $\psi$. Then, the noise probability distribution is [see Eq.~(\ref{prob:cohsq})]
\begin{equation}
    p_{\eta}(\eta,\phi)=\frac{1}{\sqrt{2\pi}\sigma_{\phi}}\exp\left[\frac{-\eta^{2}}{2\sigma_{\phi}^{2}}\right],
\end{equation}
where the variance is
\begin{equation}
    \sigma_{\phi}^{2} = \sigma_b^2 \sin^{2}\left(\phi-\psi\right) + \sigma_s^2 \cos^{2}\left(\phi-\psi\right).
\end{equation}
We see that $\sigma_{\psi} = \sigma_s$ and $\sigma_{\psi+\pi/2} = \sigma_b$ giving the expected variances in the quadratures $\hat{\eta}_{\psi}$ and $\hat{\eta}_{\psi+\pi/2}$. For a general quadrature $\hat{\eta}_{\phi}$, the variance is a convex combination of $\sigma_s^2$ and $\sigma_b^2$.

The parameter $\phi$ refers to the phase of the local oscillator used for the required optical homodyne, see Fig.~\ref{fig:Scheme}, and thus, is externally chosen for each measurement. For a given data point, one can always choose $\psi=\phi$ to get the least amount of noise, effectively yielding
\begin{equation}
    p_{\eta}(\eta,\phi) = \frac{1}{\sqrt{2\pi}\sigma_s}\exp\left[\frac{-\eta^2}{2\sigma_s^2} \right].
\end{equation}

With the calculation of $p_m$ and $p_{\eta}$, one can perform the integral in Eq.~\eqref{prob_a:conv} to find the output photon statistics. For the squeezed coherent state introduced above, 
\begin{equation} 
    \Ket{\Psi_{\rm coh}} = \hat{D}(\alpha_s) \hat{S}(r_s,\psi_s) \Ket{0},
\end{equation} 
we get a Gaussian
\begin{equation}
    p_{a,\mathrm{coh}}(a,\phi) = \frac{1}{\sqrt{2\pi}\sigma_{\rm coh}(\phi)} \exp\left[-\frac{(a-\mu_{\rm coh}(\phi))^2}{\sigma_{\rm coh}^2(\phi)} \right].
\end{equation}
The mean of the probability distribution is $\mu_{\rm coh}(\phi) = 2\sin\theta\mathrm{Re}[\alpha_s e^{-i\phi}]$, as expected from Eq.~(\ref{eq:out}). The variance is given by a weighted average of the noise variance, $\sigma_s^2$, and the magnon variance given in terms of $r_s$ and $\psi_s$,
\begin{multline} 
    \sigma_{\rm coh}^2(\phi) = \cos^2\theta \sigma_s^2 + \\ \sin^2\theta \left( e^{-2r_s} \cos^2\left(\phi-\psi_s\right) + e^{2r_s} \sin^2\left(\phi-\psi_s\right) \right).
\end{multline}
For $\theta=\pi/2$, corresponding to no noise, $p_{a,\mathrm{coh}}$ gives the probability distribution of the magnons. For $\theta=0$, we simply get squeezed vacuum noise.

For the semi-classical state $\hat{\rho}_{\rm cl}$ introduced above as a 50:50 mixture of the coherent states $\Ket{\pm \alpha_c}$, the photon's output probability distribution reads
\begin{equation}
    p_{a,\mathrm{cl}}(a,\phi) = \frac{\cosh(2az_R)}{\sqrt{2\pi}\sigma_{\rm cat}} \exp\left[-\left(\frac{a^{2} + \sigma_{\rm cat}^4 z_R^2}{2\sigma_{\rm cat}^2}\right)\right]
\end{equation}
where $\sigma_{\rm cat}^2 = \sin^2\theta + \cos^2\theta \sigma_s^2$ and $z_R+iz_I= 2\alpha_c e^{-i\phi} \sin\theta/\sigma_{\rm cat}^2$. Similar considerations as above holds for $\theta=0$ and $\theta=\pi/2$. Compare this to the case of the cat state defined above as 
\begin{equation}
    \Ket{\Psi_{\rm cat}} = \mathcal{N} \frac{\Ket{\alpha_c} + e^{i\psi_c} \Ket{-\alpha_c}}{\sqrt{2}},
\end{equation}
where $p_{a,{\rm cat}}(a,\phi) = \mathcal{N}^2 p_{a,{\rm cl}}(a,\phi) I$ with the interference term being
\begin{equation}
    I = 1 + \exp\left( \frac{-2\left|\alpha_c\right|^{2}}{1+\tan^{2}\theta/\sigma_s^2} \right) \frac{\cos\left(a z_I-\psi_c\right)}{\cosh (az_R)}. \label{pcat:int}
\end{equation}
The presence of the interference term corresponds to the coherence between the two components of the cat state. The second term in $I$ has an exponential in $\left|\alpha_c\right|^{2}$ that diminishes the interference term. If $\tan\theta/\sigma_s < |\alpha_c|$,  we get $p_{a,{\rm cat}}(a,\phi) \approx p_{a,{\rm cl}}(a,\phi)$ and information about the phase of the cat state $\psi_c$ is lost. This is an evidence of the fact that, in general, the magnon's coherence information is lost unless the SNR is high or in the presence of high input squeezing. This explains why the MLE procedure converged to a classical density matrix above, except when $\tan\theta/\sigma_s \gg 1$.

\section{Characterization of SNR} \label{sec:Numbs}

The magnon signal amplitude $\theta$ defined in Eq.~\eqref{eq:out} is a central quantity to the evaluation of the amount of magnon signal in the output field $\hat{a}_{\rm{out}}$. As mentioned above, a large coherent $\boldsymbol{e}_y$-polarized pulse of duration $T_{\rm pul}$ causes a coupling between the magnons and $\boldsymbol{e}_x$-polarized photons. For an optomagnonic waveguide, see Fig.~\ref{fig:System}, $\theta$ is given by
\begin{equation}
    \theta = \sqrt{\frac{\gamma_G \hbar}{2M_s}\frac{c A_{\rm trav}^2 T_{\rm pul}}{V_{\rm mag}}} \frac{\Theta_F + \Theta_C}{2} \frac{\Gamma l  \tau}{\left|1 - \rho^2\Gamma^2 e^{2ik_{\rm out} l}\right|} . \label{Res:theta}
\end{equation}
This formula is derived in Sec.~\ref{sec:WG_ana}, while here we discuss its physical meaning and achievable values in a YIG waveguide. Here, $\gamma_G$ is the gyromagnetic ratio, $M_s$ is the saturation magnetization, $A_{\rm trav}$ is the input light amplitude inside the magnet [in the notation of Fig.~\ref{fig:Notation}, $A_{\rm trav} = |\langle \hat{l}_{y,\mathrm{out}} \rangle|$], and $V_{\rm mag}$ is the volume of the magnet. The second term describes the magneto-optical activity, with $\Theta_F$ being the Faraday rotation per unit length and $\Theta_C$ being the Cotton-Mouton ellipticity per unit length. The optical dissipation is included as $\Gamma = e^{-\opabs l/2}$ where $\opabs$ is the optical absorption coefficient per unit length and $l$ is the length of the magnet. $\rho$ and $\tau$ are the reflectivity and transmittivity at the air-magnet interface that can be controlled by a coating or a secondary waveguide. Due to time-reversal symmetry, they must satisfy $\nY \tau^2 + \rho^2 = 1$ where $\nY$ is the refractive index of the magnet. The wave-vector of the output light $k_{\rm out} = k_{\rm in} + \omega_m/v$ is given in terms of the input optical wave-vector $k_{\rm in}$, the magnon frequency $\omega_m$, and the speed of light inside the magnet $v=c/\nY$.

The light amplitude inside the magnet must be limited to avoid excessive heating. The traveling power inside the magnet is $P_{\rm trav} = A_{\rm trav}^2 \hbar\omega_{\rm in} v$, and hence the stored optical energy is $E_{\rm stored} = P_{\rm trav} l/v$. The dissipated power is $P_{\rm diss} = \opabs v E_{\rm stored}$, giving the rate of change of the magnet's temperature
\begin{equation}
    \dot{T} = \frac{P_{\rm diss}}{C_V \mu_{\rm den} V_{\rm mag}},
\end{equation}
where $C_V$ is the specific heat and $\mu_{\rm den}$ is the mass density of the magnet. To avoid significant heating, we choose an optical amplitude $A_{\rm trav}$ such that the temperature increase during the pulse is sufficiently small $k_B \dot{T} < 0.1\hbar\omega_m/T_{\rm pul}$. This ensures a magnon number increase of $<0.1$. Note that we assumed that all of the optically dissipated power goes into the magnons, which is an overestimation as the heat can sink into the phonons and also leak away from the system.

This gives 
\begin{equation}
    \frac{c A^2_{\rm trav} T_{\rm pul}}{V_{\rm mag}} = 0.1 \frac{\omega_m \nY C_V \mu_{\rm den}}{k_B \opabs \omega_{\rm in} l}.
\end{equation}
Inserting this into the expression for $\theta$, we see that the dependencies on $T_{\rm pul}$ and $V_{\rm mag}$ drops out. This implies that we can decrease the pulse width $T_{\rm pul}$ down to as small of a value as feasible by pumping up the input power, making the measurements extremely fast. Similarly, the volume dependence cancels off because the single photon coupling scaling of $\propto 1/\sqrt{V_{\rm mag}}$ is compensated by the maximum input power scaling of $\propto \sqrt{V_{\rm mag}}$.

The reflection at the boundary, $\rho$, can be tuned by design. If both the resonance conditions are satisfied, i.e. the input frequency $\omega_{\rm in}$ and output frequency $\omega_{\rm in}+\omega_m$ are multiples of $\pi v/l$, then $\theta$ is maximized at $\rho\approx\Gamma$ for both $\rho,\Gamma\approx 1$. This is akin to impedance matching maximizing the output signal. One can always adjust $\omega_{\rm in}$ by a small amount to ensure that the input resonance condition is satisfied, i.e. $k_{\rm in} l/\pi$ is an integer. This will typically make the output photons non-resonant, so one should make sure that the linewidth is larger than the degree of off-resonance, i.e. $1-\rho^2\Gamma^2 > 2\omega_m l/v$ [see Eq.~\eqref{Res:theta}]. Concretely, we get the following dependence on the reflectivity $\rho$,
\begin{equation}
    \theta \propto \frac{\sqrt{1-\rho^2}}{\left|1 - \rho^2\Gamma^2 e^{2i\omega_ml/v}\right|},
\end{equation}
using $\nY \tau^2 = 1 - \rho^2$ as mentioned above. The maximum $\theta$ is achieved at
\begin{equation}
    \rho^2 = 1 - \frac{\sqrt{1+\Gamma^4 - 2\Gamma^2 \cos(2\omega_ml/v)}}{\Gamma^2}. \label{rho:opt}
\end{equation}
For large absorption, specifically $2\Gamma^2\cos(2\omega_ml/v) < 1$, the right hand side becomes negative implying that $\theta$ is maximized at $\rho = 0$, i.e. no reflection. In our case, typically $1-\Gamma \ll \omega_ml/v \ll 1$ giving an estimate $1 - \rho \sim \omega_ml/v$. In this off-resonance limit, we get
\begin{equation}
    \theta_{\rm off} \sim \frac{1}{8\sqrt{5}} \sqrt{\frac{\gamma_G \hbar C_V \mu_{\rm den} c}{k_B M_s}} \frac{\Theta_F+\Theta_C}{\sqrt{\opabs \omega_{\rm in}}}. \label{Sabs_approx}
\end{equation}
We see that the dependence on the magnet's length $l$ also drops off. While the interaction time increases with the length of the magnet, the output photons become more off-resonance.

The parameter $\theta$ depends on the optical and magneto-optical parameters of the material which, in turn, depend on the frequency of light propagating in the material and the temperature. As a material, we consider YIG, and consider operation at two wavelengths, visible ($\lambda_1 = \SI{550}{\nano\meter}$) and infrared ($\lambda_2 = \SI{1.5}{\micro\meter}$) at cryogenic temperatures. For YIG, the wavelength $\lambda_1$ is close to the electronic band gap giving a very high MO activity at the cost of high optical absorption. In the infrared, YIG is nearly transparent with a measurable MO activity. As MO activity is caused by energy level transitions, its temperature dependence is weak~\cite{wettling_magneto-optics_1976,Boudiar_MO,Booth_MO_Infra}, so we assume $\Theta_{F,C}^{\rm cryo} = \Theta_{F,C}^{\rm RT}$, given by $\Theta_{F,\lambda_1} = \SI{3000}{\degree\per\centi\meter}$~\cite{wettling_magneto-optics_1976}, $\Theta_{F,\lambda_2} = \SI{200}{\degree\per\centi\meter}$~\cite{Boudiar_MO,Booth_MO_Infra}, and $\Theta_C\ll\Theta_F$~\cite{wettling_magneto-optics_1976}. However, the optical absorption has a strong temperature dependence~\cite{Wood_Remeika}, typically decreasing by 2-3 orders of magnitude from room temperature to cryogenic conditions. Thus, we take as an estimate for the absorption at cryogenic temperatures $\opabs^{\rm cryo} = 10^{-2} \opabs^{\rm RT}$, where the room temperature values are~\cite{Wood_Remeika} $\opabs^{{\rm RT}, \lambda_1} = \SI{200}{\per\centi\meter}$ and $\opabs^{{\rm RT}, \lambda_2} < \SI{0.03}{\per\centi\meter}$.

For YIG, $C_V = \SI{590}{\joule\per\kilogram\per\kelvin}$ and $\mu_{\rm den} = \SI{5}{\gram\per\cubic\centi\meter}$. Besides the optical frequency and the material parameters, the only remaining free parameter is the length of the magnet. We consider magnets with lengths $l\in (1,1000)\si{\micro\meter}$ \cite{YIGWaveguide,YIGBridge} corresponding to reflection coefficients $1- \rho \in (10^{-4},10^{-1})$. Varying the length, we get $|S| \in (0.18,0.25)$ for infrared and $|S| \approx 0.02$ for visible input. These values imply that the infrared light is much more efficient for magnon tomography than the visible light, for YIG. To increase the signal further, one needs to increase the optical power and sink the heat away from the magnet, e.g. using a material with a high thermal conductivity. Another way would be to design the cavity such that both the input and the output photons are in resonance, e.g. by polarization-sensitive mirrors at the edges of the waveguide.

\begin{table}
    \centering
    \begin{tabular}{|c p{7.4cm}|}
        \hline
        \textbf{Symbol} & \textbf{Definition} \\
        \hline
        $v$ & Speed of light in the magnet \\
        $\omega_{\rm in}$, $k_{\rm in}$ & Optical input frequency and wave vector $k_{\rm in} = \omega_{\rm in}/v$ \\
        $\opabs$ & Optical absorption per unit length in the magnet \\
        $\Gamma$ & Amplitude decay in one trip $e^{-\opabs l /2}$ \\
        $\rho(\tau)$ & Reflectivity(transmittivity) at waveguide boundary \\
        $\Theta_F$ & Faraday rotation per unit length \\
        $\Theta_C$ & Cotton-Mouton ellipticity per unit length \\
        $\omega_m$, $k_m$ & Magnon frequency and corresponding optical wavevector $k_m = \omega_m/v$ \\
        $S$, $\theta$ & Complex magnon signal amplitude, $\theta =|S|$ \\ 
        $\hat{a}_{\sigma}(z)$ & Photonic field annihilation operator with polarization $\sigma$ at position $z$ \\
        $\hat{\eta}_{\rm tot}(t)$ & The effective optical noise inside the magnet \\
        $\sigma_s (\sigma_b)$ & Standard deviation in the squeezed (non-squeezed) quadrature in the output noise \\
        \hline
    \end{tabular}
    \caption{Definitions of important symbols used in the text.}
\end{table}

\section{Lossy Optomagnonic waveguide} \label{sec:WG_ana}

In this section, we model a single-mode optical waveguide (along $\boldsymbol{e}_z$) composed of a magnetic insulator, such as YIG, as illustrated in Fig.~\ref{fig:System}. We determine the output light in terms of the inputs to the waveguide in order to have access to the underlying magnon state, deriving the expression for magnon signal amplitude $\theta$ given in Eq.~\eqref{Res:theta}. 

We consider a static applied magnetic field along $\boldsymbol{e}_y$ that saturates the magnetization. In this configuration, called the Voigt configuration,  the magnetization is perpendicular to the photon propagation direction, which maximizes the BLS cross-section~\cite{OptMag_Sanch}. In the absence of the magnetization, the two linearly polarized propagating modes (with electric fields polarized along $\boldsymbol{e}_x$ and $\boldsymbol{e}_y$) are decoupled, nevertheless inelastic BLS by magnons can transfer photons from one polarization to the other. Due to a large $\boldsymbol{e}_y$-polarized input, the magnons and $\boldsymbol{e}_x$-polarized light are coupled, such that the output contains information about the magnons' state. 

The photons within the magnetic material experience decay due to optical absorption, as well as additional noise inside the material and from the thermal input from outside. This noise competes with the magnon signal generated by BLS. To mitigate the external noise, we can inject squeezed vacuum in the $\boldsymbol{e}_x$-polarization, see Fig.~\ref{fig:System}. Only $\boldsymbol{e}_x$-polarization requires squeezing because the noise in $\boldsymbol{e}_y$-polarization is too small compared to the large coherent input. The rigorous derivation of the results can be found in appendices \ref{app:Waveguide} and \ref{app:WaveguideBLS}, while in this section, we discuss the salient features of our results.

\subsection{Propagation inside an optomagnonic waveguide}

Here, we discuss the propogation of light inside the optomagnonic waveguide, including the effects of BLS. We consider only the uniform magnetic excitations, i.e., Kittel magnons, as justified shortly below. Formally, we promote the magnetization field to an operator, and describe the magnetic excitations around the saturation direction as a bosonic excitation via the Holstein-Primakoff transformation,
\begin{equation}
    \hat{M}_{z}-i\hat{M}_{x}\approx2\ZPF\hat{m},\label{def:HP-1}
\end{equation}
where we ignore terms of higher order in $\hat{m}$, and the zero point fluctuations are given by
\begin{equation} 
    \ZPF=\sqrt{\frac{\gamma_G\hbar M_{s}}{2V_{{\rm mag}}}},
\end{equation}
with $\gamma_G$ being the absolute value of the gyromagnetic ratio, $M_{s}$ the saturation magnetization, and $V_{\rm mag}$ the volume of the magnet. Within this approximation, the magnon Hamiltonian is $\hat{H}_{\rm mag}=\hbar\omega_{m}\hat{m}^{\dagger}\hat{m}$~\cite{StanPrabh}, where $\omega_{m}$ is the resonance frequency set by the external magnetic field strength and the shape anisotropy. For magnetic fields of $<\SI{200}{\milli\tesla}$, the angular frequencies of magnons remain $<2\pi\times\SI{10}{\giga\hertz}$. Given that optical frequencies typically extend into the hundreds of terahertz, the photon frequencies remain nearly unaltered in BLS, and the same holds also for the magnitude of photon momentum. Due to momentum conservation, our geometry enforces a constraint on magnon momentum, allowing it to be either zero or twice the magnitude of photon momentum. As the analysis is analogous for both scenarios, we consider only the case of zero momentum magnons as stated above.

The photon's electric field is given by $\vecOp{E} = \vecOp{\cal E} + \vecOp{\cal E}^{\dagger}$ where
\begin{equation}
   \vecOp{\cal E}(\boldsymbol{r}) = \sum_{\sigma} \int_{-\infty}^{\infty} \frac{dk}{\sqrt{2\pi}} e^{ikz} E(x,y) \boldsymbol{e}_{\sigma} \hat{a}_{\sigma}(k),
\end{equation}
with $\boldsymbol{r}=(x,y,z)$. We integrate over the wave-vector $k$ and sum over the polarization $\sigma\in\{x,y\}$. The field operators satisfy $\left[\hat{a}_{\sigma}(k),\hat{a}_{\sigma'}^{\dagger}(k')\right] = \delta_{\sigma\sigma'} \delta(k-k')$, and the electric field intensity is normalized by
\begin{equation}
    \varepsilon_{0}\varepsilon_{r}\int dxdy \left|E(x,y)\right|^2 = \frac{\hbar\left|k\right|v}{2},
\end{equation}
where $\varepsilon_0$ is the permittivity of free space, $\varepsilon_r$ is the relative permittivity of the magnet, and $v=c/\nY$ is the speed of light in the magnet with $\nY$ being the refractive index of the magnet. The above normalization ensures that the optical Hamiltonian becomes that of field harmonic oscillators,
\begin{equation}
    \hat{H}_{{\rm opt}} = \sum_{\sigma} \hbar v\int dk\left|k\right|\hat{a}_{\sigma}^{\dagger}(k)\hat{a}_{\sigma}(k).
\end{equation}

The scattering of light by magnons is modelled by the Hamiltonian~\cite{OptMag_Sanch},
\begin{multline}
    \hat{H}_{{\rm int}} = \frac{c\sqrt{\varepsilon_{r}}\varepsilon_{0}}{\omega_{{\rm opt}}}\int dV \\ \left[i\frac{\Theta_{F}}{M_{s}}\hat{\boldsymbol{M}} \cdot \left(\hat{\boldsymbol{\cal E}}^{\dagger}\times\hat{\boldsymbol{{\cal E}}}\right)+\frac{\Theta_{C}}{M_{s}^{2}}\left(\hat{\boldsymbol{M}}\cdot\hat{\boldsymbol{{\cal E}}}^{\dagger}\right)\left(\hat{\boldsymbol{M}}\cdot\hat{\boldsymbol{{\cal E}}}\right)\right],
\end{multline}
where $\Theta_F$ is the Faraday rotation per unit length and $\Theta_C$ is the Cotton-Mouton ellipticity per unit length. 

We separate the photons into forward (+) and backward (-) propagating waves, $\hat{a}_{\sigma\pm}(z)$, defined as partial Fourier transforms over positive and negative wave-vectors respectively,
\begin{equation}
    \hat{a}_{\sigma \pm}(z) = \int_0^{\infty} \frac{dk}{\sqrt{2\pi}} e^{\pm ikz} \hat{a}_{\sigma}(\pm k).
\end{equation}
The propagation of the electromagnetic waves outside the magnet is that of a free wave. Therefore, the annihilation operators satisfy $\hat{a}_{\sigma\pm}(z,t)=\hat{a}_{\sigma\pm}(z \mp ct,0)$ for both $z$ and $z\mp ct$ either $>l/2$ or $<-l/2$. Inside the magnet, we have,
\begin{multline}
    \left(\partial_{t}\pm v\partial_{z}\right)\hat{a}_{\sigma\pm}(z,t)=-i\sum_{\sigma'}\hat{{\cal M}}_{\sigma\sigma'}(t)\hat{a}_{\sigma'\pm}(z,t) \\ -\frac{\gamma}{2}\hat{a}_{\sigma\pm}(z,t)-\sqrt{\gamma}\hat{\eta}_{\sigma\pm}(z,t).\label{eq:EOM:ph}
\end{multline}
Here, the operator $\hat{\cal M}_{\sigma\sigma'}(t)=G_{\sigma'\sigma}\hat{m}(t)+G_{\sigma\sigma'}^{*}\hat{m}^{\dagger}(t)$ describes the inelastic magnon Brillouin light scattering with $G_{\sigma\sigma'}$ being the scattering amplitudes. The noise source is assumed to be a local white source $\left[\hat{\eta}_{\sigma\pm}(z,t),\hat{\eta}_{\sigma'\pm}^{\dagger}(z',t')\right]=\delta_{\sigma\sigma'}\delta(z-z')\delta(t-t')$, and the dissipation rate $\gamma=v\opabs$ is given in terms of the optical absorption coefficient $\opabs$. The scattering amplitudes depend on the direction of the saturation magnetization. For the configuration shown in Fig.~\ref{fig:System}, $G_{xx} = G_{yy} = 0$, and 
\begin{equation}
\label{eq:ConvFact}
    G_{yx}=\frac{ic}{\sqrt{\varepsilon_{r}}} \frac{{\cal M}_{\rm ZPF}}{M_s} \frac{\Theta_F + \Theta_C}{2},
\end{equation}
The amplitude $G_{xy}$ is given by replacing $\Theta_F\rightarrow -\Theta_F$ in Eq.~\eqref{eq:ConvFact}.

\subsection{Input-Output relations}

\begin{figure}
\includegraphics[width=\columnwidth]{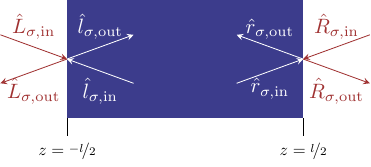}
\caption{Schematic representation of  the various boundary terms used in the main text. The external inputs for a given polarization $\sigma$ are $\hat{X}_{\sigma,{\rm in}}$, with $X = L$ or $R$, while the outputs are given by ${\rm in}\rightarrow {\rm out}$. Inside the magnet, the boundary waves are denoted by lower case letters. The total length of the magnet is $l$.
\label{fig:Notation}}
\end{figure}

The inputs and outputs to the waveguide are graphically presented in Fig.~\ref{fig:Notation}. Here, we derive the output operators in terms of the input operators. The left and right inputs are respectively written as (for $\epsilon \rightarrow 0^+$),
\begin{equation}
\begin{split}
    \hat{L}_{\sigma,{\rm in}}(t) &\equiv \hat{a}_{\sigma+}\left(\frac{-l}{2}-\epsilon,t\right), \\ 
    \hat{R}_{\sigma,{\rm in}}(t) &\equiv \hat{a}_{\sigma-}\left(\frac{l}{2}+\epsilon,t\right).
\end{split}
\end{equation}
The outputs $\hat{L}_{\sigma,{\rm out}}(t)$ and $\hat{R}_{\sigma,{\rm out}}(t)$ are defined by the replacements $\hat{a}_{\sigma\pm}\rightarrow\hat{a}_{\sigma\mp}$. The boundary amplitudes inside the magnet are indicated with the lower-case letters and found by the replacements $\epsilon\rightarrow -\epsilon$, $\hat{L}\rightarrow\hat{l}$, and $\hat{R}\rightarrow\hat{r}$.

The boundary conditions can be found by employing time-reversal symmetry and enforcing canonical commutation relations. On the left boundary,
\begin{equation}
    \begin{pmatrix} \hat{l}_{\sigma,\mathrm{out}} \\ \hat{L}_{\sigma,\mathrm{out}} \end{pmatrix} = 
    \begin{pmatrix}\nY\tau & \rho \\ -\rho & \tau \end{pmatrix}
    \begin{pmatrix}\hat{L}_{\sigma,\mathrm{in}} \\ \hat{l}_{\sigma,\mathrm{in}} \end{pmatrix}, \label{def:BDs}
\end{equation}
where the reflectivity $\rho$ and transmittivity $\tau$ satisfy $\nY\tau^{2}+\rho^{2}=1$. A similar boundary condition holds at the right interface. Here, we have ignored a small rotation of the polarization upon reflection due to the magnetization, i.e. the magneto-optical Kerr effect~\cite{Freiser_MO}.

We assume a classical input from the left that is linearly polarized along $\boldsymbol{e}_y$, such that we can replace 
\begin{equation}
    \hat{L}_{y,\mathrm{in}} \rightarrow \sqrt{ \frac{P_{\rm in}}{\hbar\omega_{\rm in}c}  } e^{-i\omega_{{\rm in}}t},
\end{equation}
where $P_{\rm in}$ is the input power, and $\omega_{\rm in}$ is the input frequency. The corresponding input amplitude inside the magnet is
\begin{equation}
    \hat{l}_{y,\mathrm{out}} \rightarrow \frac{\nY\tau}{1- \rho^2 \Gamma^2 e^{2ik_{\rm in} l}} \sqrt{ \frac{P_{\rm in}}{\hbar\omega_{\rm in}c} } e^{-i\omega_{{\rm in}}t}, \label{res:lyout}
\end{equation}
where $\Gamma=e^{-\opabs l/2}$, and $k_{\rm in} = \omega_{\rm in}/v$. The amplitude has peaks at resonances $k_{\rm in} = n\pi/l$ for an integer $n$ as expected for a Fabry-Perot cavity. 

Such an input light will scatter via magnons, resulting in an output signal at the right port polarized along  $\boldsymbol{e}_x$. We assume an optical input pulse within a time smaller than the magnon's lifetime, such that we can take $\hat{m}(t) = \hat{m}e^{-i\omega_m t}$. Here, we have ignored the back-action of $\boldsymbol{e}_x$-polarized photons on the magnons. The output amplitude is given by using the photonic equation of motion Eq.~\eqref{eq:EOM:ph} along with the boundary condition Eq.~\eqref{def:BDs},
\begin{equation}
    \hat{R}_{x,{\rm out}}(t) \approx\sum_{p=0}^{\infty} \left(\rho\Gamma\right)^{2p} \hat{\eta}_{\rm tot}^{[-2p]}(t) + \hat{\cal S}(t), \label{res:Rxout}
\end{equation}
where we identify two contributions for the output mode: the first term, proportional to the operators $\hat{\eta}_{\rm tot}^{[-2p]}(t)$, corresponds to all noise sources, while the second term, indicated by $\hat{\cal S}(t)$, encodes the signal from the magnon mode. The notation
\begin{equation}
    \hat{X}^{[-n]}(t)\equiv\hat{X}\left(t-\frac{nl}{v}\right)
\end{equation}
denotes retardation after $n$ one-way trips inside the magnet. The noise part $\hat{\eta}_{\rm tot}$ includes input noise from both the left and the right ends of the waveguide along with optical noise added inside the magnet $\hat{\eta}_{\rm mag}$,
\begin{equation}
    \hat{\eta}_{\rm tot} = -\rho\left(\hat{R}_{x,{\rm in}} - \Gamma^{2}\hat{R}_{x,{\rm in}}^{[-2]} \right) + \nY\tau^2 \Gamma \hat{L}_{x,{\rm in}}^{[-1]} + \hat{\eta}_{\rm{mag}}(t).
\end{equation}
As the output field $\hat{R}_{x,{\rm out}}$ depends on $\hat{\eta}_{\rm tot}$ retarded by an even number of one-way trips, this translates to the output being dependent on the noise from the right (left) retarded by an even (odd) number of one-way trips, i.e. $\hat{R}_{x,{\rm in}}^{[-2p]}$ and $\hat{L}_{x,{\rm in}}^{[-2p-1]}$.

The noise added inside the magnet, $\hat{\eta}_{x\pm}$ introduced in Eq.~\eqref{eq:EOM:ph}, accumulates to $\hat{\eta}_{\rm mag}$ whose correlations become
\begin{equation}
    \left\langle \hat{\eta}_{\rm mag}(t)\hat{\eta}_{\rm mag}^\dagger (t')\right\rangle =\frac{\tau^{2}}{v}\left(1+\rho^{2}\Gamma^{2}\right)\left(1-\Gamma^{2}\right)\delta\left(t-t'\right).
\end{equation}
which are zero if $\Gamma = 1$ (optical absorption $\opabs = 0$) or $\tau = 0$ (no transmission). 

The signal part is given by the expression
\begin{equation}
    \hat{\cal S}(t) = S_0 e^{-i\omega_{\rm in} t} \left( A_b \hat{m} e^{-i\omega_m t} + A_r \hat{m}^{\dagger} e^{i\omega_m t} \right)
\end{equation}
where
\begin{equation}
    S_0 = \frac{-i\Gamma l}{c} \frac{\nY \left(1-\rho^2\right) e^{ik_{\rm in} l}}{1-\rho^2\Gamma^2 e^{2i k_{\rm in}l}} \sqrt{\frac{P_{\rm in}}{\hbar\omega_{\rm in}c} }. \label{Def:S0}
\end{equation}
The blue sideband amplitude is proportional to
\begin{equation}
    A_b = \frac{G_{yx} \hat{m}e^{ik_ml/2}}{1-\rho^{2}\Gamma^{2} e^{2i(k_{\rm in} + k_m)l}},
\end{equation}
with $k_m = \omega_m/v$ being the momentum of a hypothetical photon of frequency $\omega_m$. The peaks at $\omega_{\rm in} + \omega_m = n\pi v/l$ for an integer $n$ are expected from the resonance condition. The red sideband component $A_r$ is given by a similar formula as $A_b$ with replacements $G_{yx} \rightarrow G_{xy}^*$ and $k_m\rightarrow -k_m$ \footnote{The red and the blue sidebands are asymmetric in amplitude because of interference of Faraday and Cotton-Mouton effects}. 

The input noise from outside $\{\hat{R},\hat{L}\}$ can be squeezed to reduce the noise. Adopting the notation $\hat{X}\in\{\hat{R}_{x,{\rm in}},\hat{L}_{x,{\rm in}}\}$, for a quadrature $\hat{X}e^{-i\psi} + \hat{X}^{\dagger}e^{i\psi}$ squeezed by a factor $e^{-r_{\rm in}}$ we have the following noise correlations
\begin{equation}
    \left\langle \hat{X}^{\dagger}(t) \hat{X}(t') \right\rangle = \sinh^2\frac{r_{\rm in}}{2} \delta(t-t'),
\end{equation}
for the input noise occupation and
\begin{equation}
    \left\langle \hat{X}(t) \hat{X}(t') \right\rangle = -\sinh\frac{r_{\rm in}}{2} \cosh\frac{r_{\rm in}}{2} e^{2i\psi} \delta(t-t'),
\end{equation}
for the noise coherence.

\subsection{Photon Detection} 

In the previous subsection, we found the output fields, in particular $\hat{R}_{x,\mathrm{out}}$. A homodyne measurement of this field corresponds to the weighted photonic field with weight function $p(t)$~\cite{Ulf} (see Sec.~\ref{app:Waveguide:Output}),
\begin{equation}
    \hat{a}_{\rm out} = \int dt\ p(t) \hat{R}_{x,\mathrm{out}}(t) e^{i\left(\omega_{\rm in}+\omega_{m}\right)t},
\end{equation}
where we single out the red sideband. For an input pulse of $T_{\rm pul}$, we choose $p(t)=\sqrt{c/T_{\rm pul}}$ for $|t|<T_{\rm pul}/2$ and $0$ otherwise. This ensures the commutation relation $\left[\hat{a}_{\rm out},\hat{a}_{\rm out}^{\dagger}\right] = 1$ converting the optical field into a confined bosonic mode. Assuming that the blue sideband is far away in frequency, we get by integrating Eq.~(\ref{res:Rxout})
\begin{equation}
    \hat{a}_{{\rm out}}\approx\hat{\eta}_{{\rm out} } + S\hat{m}, \label{eq:out:lin}
\end{equation}
where $S$ is given by
\begin{equation}
    S = S_0 \frac{G_{yx}}{1 - \rho^2\Gamma^2 e^{2i(k_{\rm in} + k_m)l}} \sqrt{c T_{\rm pul}}. \label{Def:S}
\end{equation}
We note that as we have ignored backaction from output photons on magnons, the above result is valid only for $|S|\ll 1$. When this condition is not satisfied, the above expression should still be valid as an order-of-magnitude estimate. For $|S|\approx \theta \ll 1$, Eq.~\eqref{eq:out:lin} is similar to Eq.~\eqref{eq:out} that was used to derive the main results in Sec.~\ref{sec:Recons}.

The noise component $\hat{\eta}_{\rm out}$ is found by integrating the noise component in Eq.~(\ref{res:Rxout}). Since the noise $\hat{\eta}_{\rm tot}$ depends on the input from outside, $\{\hat{L}_{x,\mathrm{in}},\hat{R}_{x,\mathrm{in}}\}$, that are squeezed, the integrated noise will be squeeezed as well. More concretely, the variance in the quadrature $\hat{\eta}_{\psi} = \hat{\eta}_{{\rm out} }e^{-i\psi} + \hat{\eta}_{{\rm out} }^{\dagger} e^{i\psi}$ is
\begin{equation}
    \sigma_s^2 = e^{-r_{\rm in}} + \frac{\left(1-e^{-r_{\rm in}}\right) \left(1-\Gamma^2\right) \left(1-\rho^2\right)}{1 + \rho^2\Gamma^2 - 2\rho\Gamma\cos(k_{\rm in} l + k_ml)}. \label{eq:ns:var}
\end{equation}
The squeezing is reduced compared to the waveguide input, i.e. $\sigma_s^2 > e^{-r_{\rm in}}$, due to the noise added by the magnet. Similarly, the quadrature $\hat{\eta}_{\psi + \pi/2}$ has a variance $\sigma_b^2$ given by replacing $r_{\rm in}\rightarrow -r_{\rm in}$ in Eq.~(\ref{eq:ns:var}). Finally, the two quadratures are independent, i.e. $\avg{\hat{\eta}_{\psi} \hat{\eta}_{\psi+\pi/2} + \hat{\eta}_{\psi + \pi/2} \hat{\eta}_{\psi}} = 0$. In the case of impedance matching $\rho=\Gamma$ and at resonance $(k_{\rm in} + k_m)l = 2n\pi$, we find $\sigma_s = 1$, i.e. there is no squeezing. However, to keep the coherent $\boldsymbol{e}_y$-polarized input resonant, we require $k_{\rm in}l=2n\pi$. This implies that the output photons will be slightly off-resonant as $k_ml \ll 1$. To compensate for this small detuning, the reflectivity should be slightly lower, $\rho < \Gamma$, to effectively increase the cavity linewidth up to detuning, i.e. $1-\rho \sim k_ml$. As discussed in Sec.~\ref{sec:Numbs}, this would typically imply, $1\gg 1-\rho \gg 1-\Gamma$, giving $\sigma_s^2 \approx e^{-r_{\rm in}}$.

\section{Conclusion} \label{sec:Conclusion}

We have proposed and evaluated a method for performing quantum tomography of magnon states hosted in dielectrics. In our proposal, the tomographic reconstruction is performed via optical light scattered by magnons, i.e. BLS. We discussed the reconstruction of the magnon's density matrix through the statistical analysis of the output photons using a maximum likelihood estimate (MLE). When the input light is squeezed by a factor $e^{r_{\rm in}}$ and the ratio of magnon to noise amplitude in the output is $\tan\theta$, the effective SNR becomes $\tan(\theta) e^{r_{\rm in}}$. When the effective SNR is $\ll 1$, the tomographic reconstruction fails, although the procedure still yields information about the coherent component of the magnon state, such as $\langle \hat{m} \rangle$. High effective SNRs $\tan(\theta) e^{r_{\rm in}} \gg 1$ yield good reconstruction fidelities, although meeting such a condition requires significant technological advancements.

In the realistically achievable case of $\tan(\theta) e^{r_{\rm in}} \sim 1$, we observed that regions with positive Wigner functions were accurately reproduced, while the algorithm ignored the negative regions that encode quantum features, for example the interference fringes of cat states. This issue stems from limitations in the data itself rather than the reconstruction algorithm. A substantial number of data points may be necessary to address the subtle changes in the probability distributions caused by the quantum features. We note that getting more data points than we considered here, $N=10^4$, is experimentally feasible because each optical pulse for measurement should be shorter than the magnons' lifetime ($<\SI{100}{\nano\second}$). However, it will require higher computational resources for MLE, which has a linear complexity $O(N)$. The computational efficiency can be significantly improved by using compact representations of the density matrices such as tensor networks~\cite{TensorNets}, restricted boltzmann machines~\cite{RBM,Tiunov_2020_experimentalquantumhomodyne} and other neural networks, which we leave to a future study.

We discussed the specific setup of a cylindrical optomagnonic waveguide, such that the output field of the waveguide includes a signal due to BLS by magnons. The conversion of magnons to photons is approximately independent of the geometry, so we expect our results to hold for other structures too, such as spheres~\cite{OptMag_Osada,OptMag_Zhang,OptMag_James} or optomagnonic crystals~\cite{OMag_Panta,OMag_Jasmin,SmallCav_James}. We evaluated the SNR for input light in the infrared regime, an optimal choice for our purposes since the magneto-optical activity of YIG is relatively high at such frequencies, while optical dissipation is minimal. Despite this favorable regime, we found that the SNR is still low, $<0.2$. It is indeed possible to improve the SNR by increasing the input laser power, contingent on transporting excess heat away from the magnet, e.g. by using a material with a high heat conductivity. Furthermore, we demonstrated that injecting squeezed vacuum can partially compensate for the low SNR, effectively reducing noise in the system.

While initially designed for ferromagnets, our proposed scheme can potentially be extended to antiferromagnets (AFMs). Unlike ferromagnets, AFMs exhibit a stronger BLS amplitude for processes involving two magnons than those involving a single magnon \cite{Cottam,parvini_afmoptomag}, yielding a significantly higher SNR. Nevertheless, the tomographic reconstruction has to be modified to take into account the different relationship between the output optical field and magnon operators, specifically the operator in Eq. \eqref{def:Proj} would involve two-magnon operators. Quantum tomography of magnon states in antiferromagnets can be used, for example, to reconstruct intrinsic magnon squeezed state in such materials \cite{Kamra_2019_antiferromagneticmagnons}.

\section{Acknowledgements}

VASVB acknowledges financial support from the \textit{Contrat Triennal 2021-2023 Strasbourg Capitale Europeenne}. SS and SVK acknowledges funding from the \textit{Bundesministerium f\"{u}r Bildung und Forschung} under the project \textit{QECHQS} (Grant No. 16KIS1590K).

\section{Code Availability}
The plots in the paper can be reproduced using the codes available at \cite{Codes}.

\appendix

\section{Maximum Likelihood} \label{app:ML}

Given a set of data $(a_{i},\phi_{i})$ for $i\in\{1,\dots,N\}$,
we want to find the magnon density matrix that maximizes the probability
of observing the set, i.e. that maximizes 
\begin{equation}
P=\prod_{i=1}^{N}p(a_{i},\phi_{i},\hat{\rho}),
\end{equation}
where 
\begin{equation}
    p(a,\phi,\hat{\rho})={\rm Tr}\left[\hat{\rho}\hat{P}(a,\phi)\right],
\end{equation}
with $\hat{P}(a,\phi)$ defined in Eq. (\ref{def:Proj}). To ensure hermiticity and positivity, we write $\hat{\rho}=\hat{T}^{\dagger}\hat{T}$. The trace constraint is enforced via a Lagrange multiplier, giving the optimization function,
\begin{equation}
{\cal L}[\hat{\rho}]=-\sum_{i=1}^{N}\log{\rm Tr}\left[\hat{T}^{\dagger}\hat{T}\hat{P}_{i}\right]-\lambda\left({\rm Tr}\left[\hat{T}^{\dagger}\hat{T}\right]-1\right),
\end{equation}
where $\hat{P}_{i}\equiv\hat{P}\left(a_{i},\phi_{i}\right)$ and $\lambda$ is a Lagrange multiplier. Varying
w.r.t. $\hat{T}$,
\begin{equation}
 0 = \sum_{i} \frac{ {\rm Tr} \left[ \left( \hat{T}^{\dagger} \delta\hat{T} + \delta\hat{T}^{\dagger} \hat{T} \right) \hat{P}_{i}\right] }{ {\rm Tr}\left[\hat{T}^{\dagger} \hat{T} \hat{P}_{i}\right]} - \lambda{\rm Tr}\left[\hat{T}^{\dagger} \delta\hat{T} + \delta\hat{T}^{\dagger} \hat{T} \right].
\end{equation}
As this holds for all arbitrary variations, we can replace $\delta\hat{T}\rightarrow i\delta\hat{T}$ and subtract the two results, in order to get
\begin{equation}
\sum_{i=1}^{N}\frac{\hat{P}_{i}\hat{T}^{\dagger}}{{\rm Tr}\left[\hat{T}^{\dagger}\hat{T}\hat{P}_{i}\right]}=\lambda\hat{T}^{\dagger}.
\end{equation}
Multiplying both sides by $\hat{T}$, and taking the trace gives $\lambda=N$, leading to $\hat{Z}(\hat{\rho})\hat{\rho}=\hat{\rho}$, where
\begin{equation}
    \hat{Z}(\hat\rho) = \frac{1}{N}\sum_{i=1}^{N}\frac{\hat{P}_i}{p_i},
\end{equation}
where $p_i \equiv \mathrm{Tr}[\hat{\rho} \hat{P}_i]$ is the probability of observing the data point $(a_i,\phi_i)$. Because of hermiticity of $\hat{\rho}$ and $\hat{P}_i$, we also have $\hat{\rho}\hat{Z}(\hat{\rho})=\hat{\rho}$
and therefore, $\hat{\rho}=\hat{F}(\hat{\rho})$ with
\begin{equation}
\hat{F}(\hat{\rho}) = \frac{\hat{Z}(\hat{\rho})\hat{\rho} + \hat{\rho}\hat{Z}(\hat{\rho})}{2}.
\end{equation}
The superoperator $\hat{F}$ is well-defined in the space of density matrices, i.e. for a valid $\hat{\rho}$, $\hat{F}(\hat{\rho})$ is also a valid density matrix.

We choose an initial guess $\hat{\rho}_{0}\propto\sum_{i=1}^{N}\hat{P}(a_{i},\phi_{i})$, normalized to unity trace, and find the sequence $\hat{\rho}_{i}=\hat{F}\left[\hat{\rho}_{i-1}\right]$. To show that this series converges, consider the norm square of $\hat{F}$,
\begin{equation}
    \mathrm{Tr}\left[\hat{F}[\hat{\rho}]^2\right] = \frac{1}{2N^{2}}\sum_{ij}{\rm Tr}\left[\frac{\hat{P}_i \hat{\rho} \hat{P}_j \hat{\rho} + \hat{\rho}^2 \hat{P}_i \hat{P}_j}{p_i p_j}\right].
\end{equation}
The norm can diverge if $p_i \ll 1$ but such a pair of $(a_i,\phi_i)$ has a very small probability of occurrence. To get an estimate of the norm for a given data, we can replace the summation by a probabilistic integral
\begin{equation}
    \frac{1}{N}\sum_i \rightarrow \int da_i d\phi_i p_i.
\end{equation}
Thus, we get
\begin{equation}
    \mathrm{Tr}\left[\hat{F}[\hat{\rho}]^2\right] \approx  \frac{1}{2}\int da_ida_j d\phi_id\phi_j {\rm Tr}\left[\hat{P}_i \hat{\rho} \hat{P}_j \hat{\rho} + \hat{\rho}^2 \hat{P}_i \hat{P}_j\right].
\end{equation}
Assuming the completeness of the projection operators, $\mathrm{Tr}\left[\hat{F}[\hat{\rho}]^2\right] \le \lambda^2$ where $\lambda$ is the highest eigenvalue of $\hat{\rho}$. As $\mathrm{Tr}[\hat{\rho}^2] = \sum_i \lambda_i^2$, where $\lambda_i$ are the eigenvalues of $\rho$, we conclude that $\mathrm{Tr}\left[\hat{F}[\hat{\rho}]^2\right] \le \mathrm{Tr}[\hat{\rho}^2]$, i.e. $\hat{F}$ is a contraction in the space of density matrices. Thus, the Banach fixed point theorem guarantees the convergence of the above recursion.

\section{Probability distributions} \label{app:StateGallery}

In this section, we discuss the states we considered for evaluation and the relevant probability distributions. The displacement operator is
\begin{equation}
    \hat{D}(\alpha) = \exp\left( \alpha \hat{m}^{\dagger} - \alpha^*\hat{m} \right).
\end{equation}
The squeezing operator is 
\begin{equation}
    \hat{S}(r,\psi) = \exp\left[\frac{r}{2}\left(e^{-2i\psi}\hat{m}^{2}-e^{2i\psi}\hat{m}^{\dagger,2}\right)\right].
\end{equation}
The magnon probability distributions are calculated using the defining relation,
\begin{equation}
    \left\langle \hat{f}(\hat{m}_{\phi}) \right\rangle = \int dm\  f(m) p_m(m,\phi),
\end{equation}
where $f$ is an arbitrary function, $\hat{m}_{\phi} = \hat{m}e^{-i\phi} + \hat{m}^{\dagger} e^{i\phi}$, and the average is calculated using the magnon density matrix $\hat{\rho}_m$. One can invert the above as
\begin{equation}
    p_m(m,\phi) = \left\langle \hat{\delta}(m-\hat{m}_{\phi}) \right\rangle. \label{def:pm:delta}
\end{equation}
However, such an inversion would involve a singular $\delta$-distribution that can make it difficult to keep the analysis rigorous. However, one can write these also in terms of the displacement operators using the relation
\begin{equation}
    \left\langle \hat{D}\left(i\beta e^{i\phi}\right) \right\rangle = \left\langle e^{i\beta \hat{m}_{\phi}} \right\rangle = \int dm e^{i\beta m} p_m(m,\phi)
\end{equation}
that follows from the definition of $p_m$. Inverting the above Fourier transform, we get
\begin{equation}
    p_m(m,\phi) = \int \frac{d\beta}{2\pi} \mathrm{Tr}\left[\hat{\rho} e^{i\beta(m - \hat{m}_{\phi})} \right],
\end{equation}
that gives a rigorous alternative to Eq.~(\ref{def:pm:delta}). The optical probability distributions are then calculated using Eqs.~(\ref{prob_a:conv}),
\begin{multline}
    p_a(a,\phi) = \int da'\ p_{\eta} \left(\cos\theta a-\sin\theta a',\phi\right) \\ \times p_m\left(\sin\theta a+\cos\theta a',\phi\right),
\end{multline}
and Eq.~\eqref{def:noise_prob},
\begin{equation}
    p_{\eta}(\eta,\phi) = \frac{1}{\sqrt{2\pi}\sigma_s}\exp\left[\frac{-\eta^2}{2\sigma_s^2} \right]. 
\end{equation}

For the vacuum state, $\hat{\rho}_{m,\mathrm{vac}} = \Ket{0}\Bra{0}$, we get
\begin{equation}
    \Braket{0|\hat{D}(\alpha)|0} = e^{-|\alpha|^2/2}, \label{exp:Disp00}
\end{equation}
which after Fourier transforming gives
\begin{equation}
    p_{m,{\rm vac}}(m,\phi) = \frac{e^{-m^2/2}}{\sqrt{2\pi}}.
\end{equation}
If $p_{m,\hat{\rho}}$ is known for a given $\hat{\rho}$, then for a displaced density matrix $\hat{\rho}_{\alpha} = \hat{D}(\alpha) \hat{\rho} \hat{D}^{\dagger}(\alpha)$, we get using Eq.~\eqref{def:pm:delta},
\begin{equation}
    p_{m,\hat{\rho}_{\alpha}}(m,\phi) = \mathrm{Tr}\left[\hat{\rho}_m \hat{\delta}\left(m - \hat{m}_{\phi} - 2{\rm Re}\left[\alpha e^{-i\phi}\right] \right)\right],
\end{equation}
where we used the relation
\begin{equation}
    D^{\dagger}(\alpha)\hat{m} D(\alpha) = \hat{m} + \alpha.
\end{equation}
Thus,
\begin{equation}
    p_{m,\hat{\rho}_{\alpha}}(m,\phi) = p_{m,\hat{\rho}}\left(m-2{\rm Re}\left[\alpha e^{-i\phi}\right],\phi\right),
\end{equation}

Similarly for a squeezed density matrix $\hat{\rho}_{r,\psi} = \hat{S}(r,\psi)\hat{\rho}\hat{S}^{\dagger}(r,\psi)$, we find
\begin{equation}
    p_{m,\hat{\rho}_{r,\psi}}(m,\phi) = {\rm Tr}\left[\hat{\delta}\left(z^{*}\hat{m} + z \hat{m}^{\dagger} - m\right)\hat{\rho}_m \right],
\end{equation}
where $z = \cosh re^{i\phi} - e^{i(2\psi-\phi)}\sinh r$, and we used the relation
\begin{equation}
    S^{\dagger}(r,\psi) \hat{m} S(r,\psi) = \cosh r \hat{m} - e^{2i\psi} \sinh r \hat{m}^{\dagger}.
\end{equation} 
For $z = |z|e^{i\Phi}$, we get the probability distribution
\begin{equation}
    p_{m,\hat{\rho}_{r,\psi}}(m,\phi) = \frac{1}{\left|z\right|} p_{m,\hat{\rho}} \left(\frac{m}{\left|z\right|},\Phi\right).
\end{equation}

Using these relations and the expression for $p_{m,\mathrm{vac}}$, we can calculate the probability distributions of 
\begin{equation} 
    \Ket{\Psi_{\rm coh}} = \hat{D}(\alpha_s)\hat{S}(r_s,\psi_s)\Ket{0}
\end{equation}
as
\begin{equation}
    p_{m,{\rm coh}}(m,\phi) = \frac{1}{\sqrt{2\pi}\sigma_m(\phi)}\exp\left[-\frac{\left(m-2{\rm Re}\left[\alpha_s e^{-i\phi}\right]\right)^2}{2\sigma_m^2(\phi)}\right], \label{prob:cohsq}
\end{equation}
where the angle-dependent variance is
\begin{equation}
    \sigma_m^2(\phi) = e^{-2r_s}\cos^{2}\left(\phi-\psi_s\right)+e^{2r_s}\sin^{2}\left(\phi-\psi_s\right).
\end{equation}
The corresponding optical probability distribution can be found via a straightforward Gaussian integral,
\begin{equation}
    p_{a,{\rm coh}}(a,\phi) = \frac{1}{\sqrt{2\pi}\sigma_a(\phi)}\exp\left[-\frac{\left(a-2\sin\theta{\rm Re}\left[\alpha_s e^{-i\phi}\right]\right)}{2\sigma_a^2(\phi)}\right],
\end{equation}
where the optical variance is
\begin{equation}
    \sigma_a^2(\phi) = \cos^{2}\theta \sigma_s^2 + \sin^{2}\theta\sigma_m^{2}(\phi).
\end{equation}
While the mean is scaled down by a factor $\sin\theta$, the variance is mixed with the noise variance $\sigma_s^2$.

Next, we consider a classical mixture of coherent states, given by the density matrix
\begin{equation}
    \hat{\rho}_{\rm cl} = \frac{ \Ket{\alpha_c}\Bra{\alpha_c} + \Ket{-\alpha_c}\Bra{-\alpha_c} }{2}.
\end{equation}
The magnon probability density in this case is given by putting $r_s = 0$ in $p_{m,\mathrm{coh}}$ along with the following replacements,
\begin{equation}
    2p_{m,\mathrm{cl}} = \left. p_{m,\mathrm{coh}} \right|_{\alpha_s\rightarrow \alpha_c} + \left. p_{m,\mathrm{coh}} \right|_{\alpha_s\rightarrow -\alpha_c}.
\end{equation}
Explicitly,
\begin{equation}
    p_{m,{\rm cl}}(m,\phi) = \frac{1}{\sqrt{2\pi}} \exp\left[\frac{m^2+4y_R^2}{2}\right] \cosh(2ay_R),
\end{equation}
where $\alpha_ce^{-i\phi} = y_R+iy_I$. The photon probability density is also found in a similar way using $p_{a,\mathrm{coh}}$,
\begin{multline}
    p_{a,{\rm cl}}(a,\phi) = \frac{1}{\sqrt{2\pi}\sigma}\exp\left[-\left(\frac{a^{2} + 4 y_R^2 \sin^{2}\theta}{2\sigma^{2}}\right)\right] \\ \cosh\left(\frac{2ay_R\sin\theta}{\sigma^{2}}\right)
\end{multline}
where $\sigma^2 = \sin^2\theta + \cos^2\theta \sigma_s^2$.

Next, a cat state is defined as
\begin{equation}
    \Ket{\Psi_{\rm cat}} = \mathcal{N} \frac{\hat{D}(\alpha_c) + e^{i\psi_c}\hat{D}(-\alpha_c)}{\sqrt{2}} \Ket{0},
\end{equation}
where the normalization constant is
\begin{equation}
    \mathcal{N} = \frac{1}{\sqrt{1+ e^{-2 |\alpha_c|^2}\cos\psi_c}}.
\end{equation}
The average of the product of the displacement operators $\hat{D}(\alpha)$ can be found using the relations $\hat{D}(-\alpha_c) = \hat{D}^{\dagger}(\alpha_c)$,
\begin{align}
    \hat{D}(-\alpha_c)\hat{D}(\alpha)\hat{D}(\alpha_c) &= e^{2i\mathrm{Im}[\alpha \alpha_c^*]} \hat{D}(\alpha), \\
    \hat{D}(\pm \alpha_c)\hat{D}(\alpha)\hat{D}(\pm \alpha_c) &= \hat{D}(\alpha \pm 2\alpha_c ),
\end{align}
along with Eq.~(\ref{exp:Disp00}). Finally, we get the probability distribution
\begin{multline}
    p_{m,{\rm cat}}(m,\phi) = \frac{{\cal N}^{2}}{\sqrt{2\pi}} e^{-\frac{m^{2}}{2}-2R^{2}} \\ \left[\cosh\left(2ay_R\right) + \cos\left(2ay_I - \psi_c\right)\right].
\end{multline}
The optical output's probability distribution can be written as
\begin{multline}
    p^{\rm out}_{\rm cat}(a,\phi) = \mathcal{N}^2 p_{\rm cl}^{\rm out}(a,\phi) \\ \left[1 + \exp\left(-\frac{2\left|\alpha_c\right|^2}{1+\tan^2\theta/\sigma_s^2}\right) \frac{\cos\left(\frac{2ay_I\sin\theta}{\sigma^{2}} - \psi_c \right) }{\cosh \left(\frac{2ay_R\sin\theta}{\sigma^{2}}\right) } \right].
\end{multline}

\section{Photon propagation in a lossy waveguide \label{app:Waveguide}}

Consider an infinitely long single mode optical waveguide along $\boldsymbol{e}_z$. In this section, we review propagation of photons in a lossy dielectric cylinder to set the stage to study BLS in the next section. We ignore the polarization index as the propagation for both the polarizations is independent of each other. 

The quantized electric field is given by $\hat{\boldsymbol{E}}=\hat{\boldsymbol{{\cal E}}}+\hat{\boldsymbol{{\cal E}}}^{\dagger}$, where
\begin{equation}
    \hat{\boldsymbol{{\cal E}}}(\boldsymbol{r})=\sum_{\sigma} \int \frac{dk}{\sqrt{2\pi}} e^{ikz}\boldsymbol{E}_{k}(x,y) \hat{a}(k), \label{def:Ph_Quant}
\end{equation}
Here, the spatial point $\boldsymbol{r}=(x,y,z)$, $k$ gives the wave-vector in the $z$-direction, the function $\boldsymbol{E}_{k}(x,y)$ is the transverse distribution of the electric field, and the photon field annihilation operators satisfy $\left[\hat{a}(k),\hat{a}^{\dagger}(k^{\prime})\right]=\delta(k-k')$.
The electric field functions are normalized as,
\begin{equation}
    \varepsilon_{0}\varepsilon_{r}\int dxdy\boldsymbol{E}_{k}^{*}(x,y)\cdot\boldsymbol{E}_{k}(x,y)=\frac{\hbar\left|k\right|v}{2},
\end{equation}
where $\varepsilon_{r}$ is the relative permittivity and $v$ is
the speed of light inside the dielectric. The normalization is chosen to ensure that the Hamiltonian becomes that of field harmonic oscillators,
\begin{equation}
\hat{H}_{{\rm opt}}=\hbar v\int dk\left|k\right|\hat{a}^{\dagger}(k)\hat{a}(k).
\end{equation}
The equations of motion without dissipation are 
\begin{equation}
\frac{d}{dt}\hat{a}(k)=-i\left|k\right|v\hat{a}(k).
\end{equation}
This can be converted to position space by introducing the forward and backward propagating waves as $\hat{a}_{+}(z,t)=\int_{0}^{\infty}\frac{dk}{\sqrt{2\pi}}\ e^{ikz}\hat{a}(k,t)$ and $\hat{a}_{-}(z,t)=\int_{-\infty}^{0}\frac{dk}{\sqrt{2\pi}}\ e^{ikz}\hat{a}(k,t)$. Because of the integrals being over only half-spaces, their commutation relations are not as simple as their wave-vector counterparts. Nevertheless, position dependent photonic fields are useful for introducing local dissipation sources in the equations of motion,
\begin{equation}
    \left(\partial_{t}\pm v\partial_{z}\right)\hat{a}_{\pm}(z,t)=-\frac{\gamma}{2}\hat{a}_{\pm}(z,t)-\sqrt{\gamma}\hat{\eta}_{\pm}(z,t).
\end{equation}
Here, the noise source is assumed to be a local white source $\left[\hat{\eta}_{\pm}(z,t),\hat{\eta}_{\pm}^{\dagger}(z',t')\right]=\delta(z-z')\delta(t-t')$ with other commutators being zero, and the dissipation rate $\gamma=v\opabs$ with $\opabs$ being the optical absorption coefficient. The general solution of the above differential equation is,
\begin{multline}
    \hat{a}_{\pm}\left(z,t\right) = e^{\opabs \left( \mp z - l/2 \right)/2} \hat{a}_{\pm}\left(\mp\frac{l}{2},t + \frac{\mp z - l/2}{v}\right) \mp \\ \sqrt{v \opabs} \int_{\mp l/2}^{z} \frac{d\tilde{z}}{v} e^{\mp\opabs\left(z-\tilde{z}\right)/2} \hat{\eta}_{\pm}\left( \tilde{z} , t\mp\frac{z-\tilde{z}}{v} \right). \label{eq:Ph_bulk_prop}
\end{multline}
This gives $\hat{a}_+(z,t)$ and $\hat{a}_-(z,t)$ in terms of the retarded boundary terms $\hat{a}_+(-l/2,t)$ and $\hat{a}_-(l/2,t)$ respectively in addition to added noise.

We briefly pause to check the commutation relations of photons in bulk. Putting $l\rightarrow\infty$, the first term disappears because of the exponential and the noise terms can be simplified to
\begin{equation}
    \left[\hat{a}_{\pm}(z,t),\hat{a}_{\pm}^{\dagger}(0,0)\right] = e^{\mp \opabs |z|/2} \delta\left(z \mp vt\right). \label{eq:Comm_Ph}
\end{equation}
This can be understood as photons traveling with speed $v$ and suffering an amplitude decay per unit length of $\opabs/2$. 

\subsection{Boundary conditions}

The propagation in an infinite waveguide, given by Eq.~(\ref{eq:Ph_bulk_prop}), should be supplemented by boundary conditions in the case of a finite waveguide. For simplicity, we assume that the small magneto-optical effects do not affect the boundary implying that the two polarizations can be considered independent of each other. Consider a semi-infinite magnet in the space $z>0$ with the medium on the left, $z<0$, being air. For sufficiently small $z$, the photons on the left satisfy
\begin{equation}
    z<0: \hat{a}_{\pm}(z,t)=\hat{a}_{\pm}\left(0,t\mp\frac{z}{c}\right),
\end{equation}
while the ones in the right
\begin{equation}
    z>0: \hat{a}_{\pm}(z,t)=\hat{a}_{\pm}\left(0,t\mp\frac{z}{v}\right),
\end{equation}
ignoring dissipation. We define
\begin{equation}
    \hat{l}_{\pm}(t)=\hat{a}_{\pm}(0^{-},t),\ \hat{L}_{\pm}(t)=\hat{a}_{\pm}(0^{+},t).
\end{equation}
Using Eq.~(\ref{eq:Comm_Ph}), we get
\begin{align}
    \left[\hat{L}_{\pm}(t),\hat{L}_{\pm}^{\dagger}(t')\right] &= \frac{1}{c}\delta(t-t'),\\ \left[\hat{l}_{\pm}(t),\hat{l}_{\pm}^{\dagger}(t')\right] &= \frac{1}{v}\delta(t-t').
\end{align}
We assume a general scattering matrix between inputs $\{\hat{L}_{+},\hat{l}_{-}\}$ and outputs $\{\hat{L}_{-},\hat{l}_{+}\}$,
\begin{equation}
    \begin{pmatrix} \hat{l}_{+} \\ \hat{L}_{-}\end{pmatrix} = 
    \begin{pmatrix}
        \tau_{1} & \rho_{1}\\
        \rho_{2} & \tau_{2}
    \end{pmatrix}
    \begin{pmatrix} \hat{L}_{+} \\ \hat{l}_{-} \end{pmatrix}. \label{eq:BD_basic}
\end{equation}
Correct commutation relations follow if 
\begin{equation}
    \nY = \tau_{1}^{2} + \nY\rho_{1}^{2},\ 1 = \rho_{2}^{2} + \nY\tau_{2}^{2},
\end{equation}
where $\nY$ is the refractive index of the waveguide. Ignoring the small effect of the magnetization, we assume time-reversal symmetry giving
\begin{equation}
    \begin{pmatrix}\hat{l}_{-}\\\hat{L}_{+}
    \end{pmatrix} =
    \begin{pmatrix}
        \tau_{1} & \rho_{1}\\
        \rho_{2} & \tau_{2}
    \end{pmatrix}
    \begin{pmatrix} \hat{L}_{-} \\ \hat{l}_{+} \end{pmatrix}. \label{eq:BD_TRS}
\end{equation}
Eqs. (\ref{eq:BD_basic}-\ref{eq:BD_TRS}) can be satisfied by two parameters, $\rho$ and $\tau$,
\begin{equation}
    \begin{pmatrix}\hat{l}_{+}\\ \hat{L}_{-} \end{pmatrix}=\begin{pmatrix}
        \nY \tau & \rho\\
        -\rho & \tau
    \end{pmatrix}
    \begin{pmatrix}\hat{L}_{+}\\\hat{l}_{-}\end{pmatrix},
\end{equation}
satisfying $\nY\tau^{2}+\rho^{2}=1$.

\subsection{Photon amplitude}

Now, consider a finite magnet as in Fig. \ref{fig:Notation} with sides at $z=\pm l/2$. We want to find the photon amplitude inside the magnet in terms of the inputs 
\begin{equation} \begin{split}
    \hat{L}_{{\rm in}}(t) &= \hat{a}_{+}\left(-\frac{l}{2}-\varepsilon,t\right),\\ 
    \hat{R}_{{\rm in}}(t) &= \hat{a}_{-}\left(\frac{l}{2}+\varepsilon,t\right), \label{Defs:input}
\end{split} \end{equation}
for a negligible $\varepsilon>0$. Similarly, one can define $\hat{L}_{\rm out}$ and $\hat{R}_{\rm out}$ by interchanging $\hat{a}_+\leftrightarrow\hat{a}_-$. The boundary operators inside the magnet are defined analogously, e.g. 
\begin{equation}
    \hat{l}_{\rm in} = \hat{a}_-\left(\frac{-l}{2} + \varepsilon, t\right), \label{Defs:InMag}
\end{equation}
and so on. The outside operators are related to inside operators via the boundary conditions,
\begin{equation}
    \begin{pmatrix} \hat{l}_{\rm out} \\ \hat{L}_{\rm out} \end{pmatrix} = 
    \begin{pmatrix}
       \nY \tau & \rho\\
        -\rho & \tau
    \end{pmatrix}
    \begin{pmatrix}\hat{L}_{\rm in}\\ \hat{l}_{\rm in} \end{pmatrix}, \label{Photon:BDs}
\end{equation}
which holds also with $\hat{L}\rightarrow \hat{R}$ and $\hat{l}\rightarrow \hat{r}$. The inside operators are related via the equations of motion, Eq.
(\ref{eq:Ph_bulk_prop}), 
\begin{equation}
    \begin{split}
    \hat{r}_{\rm in} &= \Gamma \hat{l}_{\rm out}^{\left[-1\right]} + \hat{\eta}_{+, {\rm{I}}}, \\
    \hat{l}_{\rm in} &= \Gamma \hat{r}_{\rm out}^{\left[-1\right]} + \hat{\eta}_{-,{\rm{I}}}.\label{eq:Ph_In_Mag}
    \end{split}
\end{equation}
Here, $\Gamma=e^{-\opabs l/2}$ gives the decay of the signal. We introduce a notation for retarded waves
\begin{equation}
    \hat{f}^{\left[-n\right]}\equiv\hat{f}\left(t-\frac{nl}{v}\right).
\end{equation}
The integrated noises are given by 
\begin{multline}
    \hat{\eta}_{\pm, {\rm{I}}}(t)=-\sqrt{v \opabs} \\ \int_{-l/2}^{l/2}\frac{d\tilde{\xi}}{v}e^{-\opabs\left(l/2\mp\tilde{\xi}\right)/2}\hat{\eta}_{\pm}\left(\tilde{\xi},t-\frac{l/2\mp\tilde{\xi}}{v}\right).
\end{multline}
This gives the variance,
\begin{equation}
\left\langle \hat{\eta}_{\pm, {\rm{I}}}(t) \hat{\eta}_{\pm, {\rm{I}}}^\dagger (t^\prime)\right\rangle =\frac{1-\Gamma^{2}}{v}\delta\left(t-t'\right).
\end{equation}
The forward wave equation in Eq. (\ref{eq:Ph_In_Mag}) can be interpreted as the photons from the left boundary, $\hat{l}_{\rm out}$, traveling through with a delay of $l/v$, a decay of $\Gamma$, and added noise $\hat{N}_{+}$. Similar interpretation holds for the backward moving wave. 

The resulting set of equations can be solved to get a recursive relation,
\begin{equation}
    \hat{l}_{\rm out} = \rho^{2}\Gamma^{2}\hat{l}_{\rm out}^{\left[-2\right]} + \hat{F}_{L},
\end{equation}
with the `force'
\begin{equation}
    \hat{F}_{L} = \nY\tau\left(\hat{L}_{{\rm in}}+\rho\Gamma\hat{R}_{{\rm in}}^{\left[-1\right]}\right)+\rho\left(\hat{N}_{-}+\rho\Gamma\hat{\eta}_{+, {\rm I}}^{\left[-1\right]}\right).
\end{equation}
The above recursion is solved by 
\begin{equation}
    \hat{l}_{\rm out}=\sum_{p=0}^{\infty}\left(\rho\Gamma\right)^{2p}\hat{F}_{L}^{[-2p]}.
\end{equation}
Similarly, we get
\begin{equation}
    \hat{r}_{\rm out}=\sum_{p=0}^{\infty}\left(\rho\Gamma\right)^{2p}\hat{F}_{R}^{[-2p]},
\end{equation}
where
\begin{equation}
    \hat{F}_{R} = \nY \tau\left(\hat{R}_{{\rm in}}+\rho\Gamma\hat{L}_{{\rm in}}^{\left[-1\right]}\right)+\rho\left(\hat{\eta}_{+, {\rm I}}+\rho\Gamma\hat{\eta}_{-, {\rm I}}^{\left[-1\right]}\right).
\end{equation}

\subsection{Output} \label{app:Waveguide:Output}
Given the photon amplitudes inside the magnet, the outputs are simply given by the boundary conditions in Eqs.~(\ref{Photon:BDs}). For concreteness, we consider the right side boundary with output $\hat{R}_{\rm out}(t)$. The output wave is collected and filtered into a confined photon operator as~\cite{Ulf}
\begin{equation}
    \hat{a}_{\rm out} = \int dt\ \tilde{p}(t) e^{i\omega_{\rm obs} t} \hat{a}_+(z_{\rm det},t),
\end{equation}
where $z_{\rm det} > l/2$ is the detector position, $\omega_{\rm obs}$ is the center frequency of the filter and $\tilde{p}(t) > 0$ is the instantaneous collection efficiency. As the evolution outside the magnet is that of a travelling wave, we get
\begin{equation}
    \hat{a}_+(z_{\rm det},t) = \hat{R}_{\rm out} \left(t-\frac{z_{\rm det}-l/2}{c}\right).
\end{equation}
giving
\begin{equation}
    \hat{a}_{\rm out} = \int dt\ p(t) e^{i\omega_{\rm obs} t} \hat{R}_{\rm out}(t).
\end{equation}
where
\begin{equation}
    p(t) = \tilde{p}\left(t + \frac{z_{\rm det}-l/2}{c}\right),
\end{equation}
and we ignored a constant phase shift. For the correct commutation relations $\left[\hat{a}_{\rm out},\hat{a}_{\rm out}^{\dagger}\right] = 1$, the function $p(t)$ must satisfy
\begin{equation}
    \int \frac{dt}{c} p^2(t) = 1. \label{p(t):norm}
\end{equation}
Typically, we take $p(t)$ to be a constant in a time window and zero outside. Similar terms can be defined with $\hat{R}\rightarrow \hat{L}$. 

\emph{Vacuum:} For a vacuum input 
\begin{equation}
\avg{\hat{X}_{{\rm in}}(t)\hat{X}_{{\rm in}}^{\dagger}(t')}=\frac{1}{c}\delta\left(t-t'\right),
\end{equation}
with $X\in\{L,R\}$. We get an intuitively appealing correlation function inside the magnet
\begin{equation}
    \left\langle \hat{l}_{\rm out}(t) \hat{l}_{\rm out}^{\dagger}(t') \right\rangle = \frac{1}{v}\sum_{\mu=-\infty}^{\infty} \left(\rho\Gamma\right)^{2\left|\mu\right|} \delta\left(t-t'-\frac{2\mu l}{v}\right).
\end{equation}
For infinite dissipation $\Gamma\rightarrow0$ or no reflection $\rho\rightarrow0$, the correlation function is that of a free wave with no retardation. Otherwise, the waves at time separation $2rl/v$ are correlated with a decrease in correlation $(\rho\Gamma)^{2\left|r\right|}$ corresponding to $2r$ reflections and dissipation during one way trips.

For the output fields $\hat{X}_{{\rm out}}(t)$, we get a vacuum state
\begin{equation}
    \left\langle \hat{X}_{{\rm out}}(t)\hat{X}_{{\rm out}}^{\dagger}(t')\right\rangle =\frac{\delta\left(t-t'\right)}{c},
\end{equation}
as expected in thermal equilibrium. This gives $\hat{a}_{\rm out}$ in vacuum state.

\emph{Squeezed Vacuum:} For a squeezed input, the correlations can be written in terms of the quadratures: $\hat{X}_{s,\mathrm{in}} = \hat{X}_{\rm in} e^{-i\psi} + \hat{X}_{\rm in}^{\dagger} e^{i\psi}$ and $-i\hat{X}_{b,\mathrm{in}} = \hat{X}_{\rm in} e^{-i\psi} - \hat{X}_{\rm in}^{\dagger} e^{i\psi}$ for $X\in\{L,R\}$ and the squeezing direction given by $\psi$. For a squeezing factor of $r$,
\begin{align*}
    \avg{\hat{X}_{s,\mathrm{in}}(t) \hat{X}_{s,\mathrm{in}}(t')} = &e^{-r} \frac{\delta(t-t')}{c} \\
    \avg{\hat{X}_{b,\mathrm{in}}(t) \hat{X}_{b,\mathrm{in}}(t')} = &e^{r} \frac{\delta(t-t')}{c} \\
    \avg{\hat{X}_{s,\mathrm{in}}(t)\hat{X}_{b,\mathrm{in}}(t') + \hat{X}_{b,\mathrm{in}}(t)\hat{X}_{s,\mathrm{in}}(t')} =& 0.
\end{align*}
Then, we can calculate the output noise statistics also in terms of similar quadratures,
\begin{multline*}
    c\left\langle \hat{X}_{s,\mathrm{out}}(t)\hat{X}_{s,\mathrm{out}}(t')\right\rangle = e^{-r}\delta(t-t') + \\ \sum_{\mu=-\infty}^{\infty} \frac{ \left(\rho\Gamma\right)^{2\left|\mu\right|} \nY\tau^{2} }{1-\rho^{2}\Gamma^{2}} \left(1-e^{-r}\right) \left(1-\Gamma^{2}\right)\delta\left(t-t'-\frac{2\mu l}{v}\right).
\end{multline*}
For $r\ne0$, the output is colored. For the collected photons,
\begin{multline*}
    \left\langle \hat{a}_{{\rm out},s}^{2}\right\rangle = e^{-r} + \left(1-e^{-r}\right) \left(1-\rho^{2}\right) \left(1-\Gamma^{2}\right) \\ \times \sum_{\mu=-\infty}^{\infty}\frac{ \left(\rho\Gamma\right)^{2\left|\mu\right|} }{ 1-\rho^{2}\Gamma^{2} } e^{2i\mu \omega_{\rm obs} l/v} \int \frac{dt}{c} p\left(t - \frac{2\mu l}{v}\right) p(t).
\end{multline*}
If $p(t)$ is nearly constant over a width $\tau$, we can approximate $p(t-2\mu l/v) \approx p(t)$ for $\mu < 2v\tau/l$. Additionally, If $(\rho\Gamma)^{v\tau/l} \ll 1$, then the contribution of $\mu > 2v\tau/l$ would be negligible, and we can use the normalization of $p(t)$, Eq.~(\ref{p(t):norm}), to get the result in the main text, Eq.~(\ref{eq:ns:var}). For a $\tau>\SI{100}{\nano\second}$ and $l<\SI{100}{\micro\meter}$, we have $v\tau/l > 1000$, which should be sufficient for this condition to hold.

\emph{Classical input:} We consider a classical input from the left, i.e. 
\begin{equation} 
    \avg{\hat{L}_{\rm in}(t)} = \sqrt{\frac{P_{\rm in}}{\hbar \omega_{\rm in}}} e^{-i\omega_{\rm in} t}.
\end{equation}
The field inside the magnet can be found using
\begin{equation}
    \avg{\hat{l}_{\rm out}(t)} = \frac{\nY \tau}{1 - \rho^2\Gamma^2 e^{2il\omega_{\rm in}/v}} \sqrt{\frac{P_{\rm in}}{\hbar \omega_{\rm in}}} e^{-i\omega_{\rm in} t} .
\end{equation}
The output field towards the right is
\begin{equation}
    \avg{\hat{R}_{\rm out}(t)} = \frac{\nY \tau^2 \Gamma e^{il\omega_{\rm in}/v}}{1 - \rho^2\Gamma^2 e^{2il\omega_{\rm in}/v}} \sqrt{\frac{P_{\rm in}}{\hbar \omega_{\rm in}}} e^{-i\omega_{\rm in} t}.
\end{equation}
This gives a coherent $\hat{a}_{\rm out}$ with a significant amplitude only if $\omega_{\rm obs} \approx \omega_{\rm in}$.

\section{Optomagnonic waveguide \label{app:WaveguideBLS}}

In this appendix, we discuss the effect of BLS on a large $\boldsymbol{e}_y$-polarized input, giving a scattered output in $\boldsymbol{e}_x$-polarization. We assume that the ground state magnetization is along $\boldsymbol{e}_y$-direction.

We assume that only the uniform mode is excited and quantize it by the Holstein-Primakoff transformation,
\begin{equation}
    \hat{M}_{z}-i\hat{M}_{x}\approx2\ZPF\hat{m},
\end{equation}
where we ignore terms higher order in $\hat{m}$ and the zero point fluctuations are given by,
\begin{equation}
    \ZPF=\sqrt{\frac{\gamma_G \hbar M_{s}}{2V_{{\rm mag}}}},
\end{equation}
with $\gamma_G$ being the absolute value of the gyromagnetic ratio, $M_{s}$ being the saturation magnetization, and $V_{{\rm mag}}$ being the volume of the magnet. The generalization to all magnon modes is straightforward. The magnon Hamiltonian is $\hat{H}_{{\rm mag}}=\hbar\omega_{m}\hat{m}^{\dagger}\hat{m}$, where $\omega_{m}$ is the resonance frequency.

The interaction Hamiltonian is that of scattering of light by the magnetization, 
\begin{multline}
    \hat{H}_{{\rm int}} = \frac{c\sqrt{\varepsilon_{r}}\varepsilon_{0}}{\omega_{{\rm opt}}}\int dV \\ \left[i\frac{\Theta_{F}}{M_{s}}\hat{\boldsymbol{M}} \cdot \left(\hat{\boldsymbol{\cal E}}^{\dagger}\times\hat{\boldsymbol{{\cal E}}}\right)+\frac{\Theta_{C}}{M_{s}^{2}}\left(\hat{\boldsymbol{M}}\cdot\hat{\boldsymbol{{\cal E}}}^{\dagger}\right)\left(\hat{\boldsymbol{M}}\cdot\hat{\boldsymbol{{\cal E}}}\right)\right],
\end{multline}
where $\hat{\boldsymbol{\cal E}}$ is given by a generalization of Eq.~(\ref{def:Ph_Quant}), \begin{equation}
    \hat{\boldsymbol{{\cal E}}}(\boldsymbol{r})=\sum_{\sigma}\int\frac{dk}{\sqrt{2\pi}}e^{ikz}\boldsymbol{E}_{\sigma k}(x,y)\hat{a}_{\sigma}(k),
\end{equation}
Using the quantization, we get the interaction Hamiltonian
\begin{equation}
    \hat{H}_{{\rm int}} = \hbar\hat{m}\sum_{\sigma,\sigma'}\int\frac{dkdk'}{2\pi}\ \hat{a}_{\sigma}(k)\hat{a}_{\sigma'}^{\dagger}(k')g_{\sigma\sigma'}(k,k')+h.c.
\end{equation}
The coupling is
\begin{multline*}
    \hbar g_{\sigma\sigma'}(k,k') = \frac{c\sqrt{\varepsilon_{r}}}{\omega_{{\rm opt}}} \frac{{\cal M}_{{\rm ZPF}}}{M_{s}} \varepsilon_{0} \int dV\ e^{i\left(k-k'\right)z} \\ \bigg[i\Theta_{F}\boldsymbol{E}_{\sigma'k'}^{*}(x,y)\times\boldsymbol{E}_{\sigma k}(x,y) + \\ \Theta_{C}\left(\boldsymbol{e}_y\cdot\boldsymbol{E}_{\sigma'k'}^{*}(x,y)\boldsymbol{E}_{\sigma k}(x,y)+\boldsymbol{e}_y\cdot\boldsymbol{E}_{\sigma k}(x,y)\boldsymbol{E}_{\sigma'k'}^{*}(x,y)\right) \bigg]_{+},
\end{multline*}
where $A_{+}=A_{z}+iA_{x}$. The final Hamiltonian is $\hat{H}=\hat{H}_{{\rm opt}}+\hat{H}_{{\rm mag}}+\hat{H}_{{\rm int}}$.

We can follow a similar procedure as App.~\ref{app:Waveguide}, to define travelling waves $\hat{a}_{\sigma\pm}$. Their equations of motion should be supplemented with the BLS interaction. For the uniform magnon mode, we expect no backscattering, so the waves in opposite direction move independent of each other via the equations of motion (dissipation to be added later),
\begin{multline}
    \left(\partial_{t}\pm v\partial_{z}\right)\hat{a}_{\sigma\pm}(z,t) = -i\sum_{\sigma'}\int\frac{dz'}{2\pi} \\ \left[g_{\sigma'\sigma\pm}(z',z)\hat{m}(t)+g_{\sigma\sigma'\pm}^{*}(z,z')\hat{m}^{\dagger}(t)\right]\hat{a}_{\sigma'\pm}(z',t),
\end{multline}
where the position dependent couplings $g_{\sigma\sigma'\pm}$ are defined as Fourier transforms,
\begin{equation}
    g_{\sigma\sigma'\pm}(z,z') = \iint_{I_{\pm}}\frac{dkdk'}{2\pi}\ g_{\sigma\sigma'}(k,k')e^{-ikz+ik'z'},
\end{equation}
with the intervals $I_{+}=(0,\infty)$ and $I_{-}=(-\infty,0)$.

BLS occurs between photons whose wave-vector is separated by $\omega_{m}/v \ll k_{\rm opt}$, $k_{\rm opt}$ being a nominal optical wave-vector. For frequencies not too close to the threshold of the waveguide, $\boldsymbol{E}_{\sigma k}(x,y)$ has a very weak dependence on $k$, such that we can replace $\boldsymbol{E}_{\sigma k}\approx\boldsymbol{E}_{\sigma k_{{\rm opt}}}$. This gives a local relation, $g_{\sigma\sigma'\pm}(z,z') \approx 2\pi G_{\sigma\sigma'}\delta\left(z-z'\right)$, where
\begin{multline}
    G_{\sigma\sigma'} = \frac{ic\sqrt{\varepsilon_{r}}}{\hbar \omega_{{\rm opt}}} \frac{{\cal M}_{{\rm ZPF}}}{M_{s}} \varepsilon_0 \int d^{2}\boldsymbol{\rho} \bigg[\Theta_{F}\boldsymbol{E}_{\sigma'}^{*}(\boldsymbol{\rho})\times\boldsymbol{E}_{\sigma}(\boldsymbol{\rho}) + \\ \Theta_C \left( \boldsymbol{e}_y \cdot \boldsymbol{E}_{\sigma'}^{*}(\boldsymbol{\rho}) \boldsymbol{E}_{\sigma}(\boldsymbol{\rho}) + \boldsymbol{e}_y\cdot\boldsymbol{E}_{\sigma}(\boldsymbol{\rho})\boldsymbol{E}_{\sigma'}^{*}(\boldsymbol{\rho})\right)\bigg]_+ \label{Def:Gpol}
\end{multline}
suppressing the $k_{\rm opt}$-index. For typical cases of the polarization being along $\boldsymbol{e}_x$ and $\boldsymbol{e}_y$, we have $G_{xx} = G_{yy} = 0$.

Finally, we model the optical dissipation as a local source,
\begin{multline}
    \left(\partial_{t}\pm v\partial_{z}\right)\hat{a}_{\sigma\pm}(z,t)=-i\sum_{\sigma'}\hat{{\cal M}}_{\sigma\sigma'}(t)\hat{a}_{\sigma'\pm}(z,t) \\ -\frac{\gamma}{2}\hat{a}_{\sigma\pm}(z,t)-\sqrt{\gamma}\hat{\eta}_{\sigma\pm}(z,t).\label{eq:EOM_ph_z}
\end{multline}
Here, the magnon operator ${\cal M}_{\sigma\sigma'}(t)=G_{\sigma'\sigma}\hat{m}(t)+G_{\sigma\sigma'}^{*}\hat{m}^{\dagger}(t)$, and the noise source is assumed to be a local white source $\left[\hat{\eta}_{\sigma\pm}(z,t),\hat{\eta}_{\sigma'\pm}^{\dagger}(z',t')\right]=\delta_{\sigma\sigma'}\delta(z-z')\delta(t-t')$.

The above equations can be solved analytically to give,
\begin{multline}
    \hat{a}_{\sigma\pm}\left(z,t\right) = 
    e^{\mp\gamma\left(z\pm l/2\right)/2v} \hat{a}_{\sigma+} \left( \mp \frac{l}{2} , t \mp \frac{z\pm l/2}{v}\right) \mp \\ 
    \int_{\mp l/2}^z \frac{d\tilde{\xi}}{v} e^{\mp\gamma\left(z-\tilde{\xi}\right)/2v} \Bigg[\sqrt{\gamma}\hat{\eta}_{\sigma\pm}\left( \tilde{\xi} , t \mp \frac{z-\tilde{\xi}}{v} \right) + \\ i\sum_{\sigma'} \hat{{\cal M}}_{\sigma\sigma'} \left(t\mp\frac{z-\tilde{\xi}}{v}\right) \hat{a}_{\sigma'\pm}\left( \tilde{\xi} , t \mp \frac{z-\tilde{\xi}}{v} \right) \Bigg]. \label{Sol:Opt:Full}
\end{multline}
For the forward moving wave: the first term is the propogation of the amplitude from the left boundary and the second term integrates the scattered light via BLS and noise.

For $\boldsymbol{e}_y$-polarization, we assume that the input is large enough such that we can ignore any backaction from the magnons. We want to find the output in $\boldsymbol{e}_x$-polarization. The inputs and outputs are defined analogous to Eqs.~(\ref{Defs:input}) and (\ref{Defs:InMag}) but with a polarization index. The output towards the right of the magnet is,
\begin{equation}
    \hat{R}_{x,\mathrm{out}}=-\rho\hat{R}_{x,\mathrm{in}} + \tau \hat{r}_{x,\mathrm{in}}.
\end{equation}
$\hat{R}_{x,\mathrm{in}}$ is the noise from the outside (which can be squeezed). To find the amplitudes inside the magnet, we use Eq.~(\ref{Sol:Opt:Full}) to get
\begin{multline}
    \hat{r}_{x,\mathrm{in}}(t) = \Gamma\hat{l}_{x,\mathrm{out}}^{[-1]} + \hat{\eta}_{x+, {\rm I}} - i\int_{-l/2}^{l/2}\frac{d\tilde{\xi}}{v} e^{-\opabs \left(l/2-\tilde{\xi}\right)/2} \\ \hat{\cal M}_{xy} \left(t-\frac{l/2-\tilde{\xi}}{v}\right) \hat{a}_{y+}\left(\tilde{\xi},t-\frac{l/2-\tilde{\xi}}{v}\right),
\end{multline}
where we used $G_{xx} = 0$. The noise is given by
\begin{equation}
    \hat{\eta}_{x+, {\rm I}}=-\sqrt{\gamma}\int_{-l/2}^{l/2}\frac{d\tilde{\xi}}{v}e^{-\gamma\left(l/2-\tilde{\xi}\right)/2v}\hat{\eta}_{x+}\left(\tilde{\xi},t-\frac{l/2-\tilde{\xi}}{v}\right).
\end{equation}
We can treat $\hat{a}_{y+}$ classically for a large coherent input (see the previous section), 
\begin{equation}
    \hat{a}_{y+}\left(z,t\right) = 
    e^{-\gamma\left(z + l/2\right)/2v} \hat{l}_{y,\mathrm{out}} \left( t - \frac{z + l/2}{v}\right),
\end{equation} 
which can be use to simplify the expression for $\hat{r}_{x,\mathrm{in}}(t)$ to
\begin{multline}
    \hat{r}_{x,\mathrm{in}}(t) = \Gamma\hat{l}_{x,\mathrm{out}}^{[-1]} + \hat{\eta}_{x+, {\rm I}} - ie^{-\opabs l/2} \hat{l}_{y,\mathrm{out}}\left(t-\frac{l}{v}\right) \\ \times \int_{-l/2}^{l/2}\frac{d\tilde{\xi}}{v} \hat{\cal M}_{xy} \left(t-\frac{l/2-\tilde{\xi}}{v}\right).
\end{multline}
Ignoring backaction on magnons, we assume a time-dependence $\hat{m}(t) = \hat{m} e^{-i\omega_m t}$. For $\omega_ml/v \ll 1$, we find
\begin{multline}
    \hat{r}_{x,\mathrm{in}} = \Gamma\hat{l}_{x,\mathrm{out}}^{[-1]} + \hat{\eta}_{x+, {\rm I}} - i\frac{\Gamma l}{v} \\ \times \left(G_{yx}\hat{m}e^{ik_{m}\left(l/2-vt\right)} + G_{xy}^*\hat{m}^{\dagger}e^{-ik_{m}\left(l/2-vt\right)} \right) \hat{l}_{y,\mathrm{out}}^{[-1]}
\end{multline}
%\begin{equation}
%    \hat{L}_{x,M-} = \Gamma\hat{R}_{x,M-}^{[-1]} + \hat{N}_{x-} - i\frac{\Gamma l}{v} \left(G_{yx}\hat{m}e^{ik_{m}\left(l/2-vt\right)} + G_{xy}^*\hat{m}^{\dagger}e^{-ik_{m}\left(l/2-vt\right)} \right) \hat{R}_{y,M-}^{[-1]}
%\end{equation}
The above equation generalizes the result in App.~\ref{app:Waveguide} for the output photons by adding the magnon contribution. Thus, we can write analogously
\begin{equation}
    \hat{R}_{x,\mathrm{out}} = \hat{R}_{x,\mathrm{out}}^{(0)} + \hat{R}_{x,\mathrm{out}}^{(M)}.
\end{equation}
The first term is a squeezed vacuum, as discussed in subsection \ref{app:Waveguide:Output}. The second term is given by
\begin{multline}
    \hat{R}_{x,\mathrm{out}}^{(M)}(t) = \mathcal{S}_0 \Bigg( \frac{G_{yx}\hat{m}e^{ik_m l/2} e^{-i(\omega_{\rm in}+\omega_m) t}}{1-\rho^2\Gamma^2 e^{-2ik_+l}} \\ + \frac{G_{xy}^*\hat{m}^{\dagger}e^{-ik_ml/2} e^{-i(\omega_{\rm in}-\omega_m) t}}{1-\rho^2\Gamma^2 e^{-2ik_-l}} \Bigg),
\end{multline}
where $k_{\pm} = (\omega_{\rm in} \pm \omega_m)/v$, we assumed a classical $\hat{l}_y$, and 
\begin{equation}
    \mathcal{S}_0 = -i\frac{\tau \Gamma l}{v} \frac{\nY \tau e^{i\omega_{\rm in} l/v}}{1 - \rho^2\Gamma^2 e^{2il\omega_{\rm in}/v}} \sqrt{\frac{P_{\rm in}}{\hbar \omega_{\rm in}}}
\end{equation}

\bibliography{References}

\end{document}